\documentclass{article}
\usepackage[utf8]{inputenc}
\usepackage{textgreek}
\usepackage{lscape}
\usepackage{tabularx}

\usepackage{graphicx}
\usepackage{authblk}
\usepackage{caption}
\usepackage{float}
\usepackage{array}
\usepackage{longtable}
\usepackage{booktabs}
\usepackage{ragged2e}
\usepackage{amsmath, amssymb}
\usepackage{textcomp}
\usepackage{geometry}
\usepackage{amsmath, amssymb}
\usepackage{url}
\usepackage[utf8]{inputenc}
\usepackage{cite}  

\title{\vspace{-1cm}
Complexity-Informed Causal Modeling of Neurodevelopmental Trajectories in Pediatric High-Grade Gliomas: Divergences from Neural Stem Cell Signatures}
\author[1,2,3]{Abicumaran Uthamacumaran}
\author[1,3,4,5]{Hector Zenil\thanks{Corresponding author: hector.zenil@algocyte.ai, hector.zenil@kcl.ac.uk}}

\affil[1]{\normalsize\text{ }Oxford Immune Algorithmics, Oxford University Innovation and London Institute of Healthcare Engineering, Oxford and London, U.K.}
\affil[2]{\normalsize\text{ }McGill University, Departments of Surgical and Interventional Sciences, Neurosurgical Simulation and Artificial Intelligence Learning Centre, Montreal, Canada.}
\affil[3]{\normalsize\text{ }Algorithmic Dynamics Lab, Research Departments of Biomedical Computing and Digital Twins, School of Biomedical Engineering and Imaging Sciences, King's College London, U.K.}
\affil[4]{\normalsize\text{ }King's Institute for Artificial Intelligence, King's College London, U.K.}
\affil[5]{\normalsize\text{ }Cancer Research Interest Group, The Francis Crick Institute, London, U.K.\vspace{-1cm}}
\date{}
\begin{document}

\maketitle

\begin{abstract}
Pediatric high-grade gliomas (pHGGs) are lethal evolutionary disorders with stalled developmental trajectories and disrupted differentiation hierarchies. We integrate transcriptional and algorithmic network complexity-based perturbation analysis to elucidate gene expression patterns and molecular divergence between Diffuse Midline Gliomas (DMG; previously DIPG) and glioblastoma (GBM), revealing shared developmental programs steering cell fate decision-making. Our complex systems approach supports the emerging paradigm that pHGGs are neurodevelopmental disorders with hybrid lineage identities and disrupted patterning. We identify dysregulated neurodevelopmental and morphogenetic signatures, alongside bioelectric and neurotransmitter signaling programs that alter synaptic organization, neuronal fate commitment, and phenotypic plasticity, regulating glioma phenotypic switching (cell state-transitions). Causal drivers and regulators of plasticity were predicted as both biomarkers and therapeutic targets, reinforcing the view that pediatric gliomas are disorders of cell fate decisions and collective cellular identity. Activation of neuronal differentiation programs suggests malignant fates, while blocked from terminal commitment, reflect arrested development along a neuronal lineage. Decoded plasticity signatures indicate a teleonomic bias toward neural progenitor or neuron-like identities, while synaptic transmission gene enrichment supports neuron–glioma interactions shaping fate trajectories. Using graph network-based complexity signatures, our algorithms identified interpretable plasticity markers and regulatory drivers underlying glioma aggressivity, evolvability, and decision-making. These findings advance ‘differentiation therapy’ as a systems medicine strategy, discovering plasticity regulators to reprogram malignant fates toward stable lineages, offering complex systems-based targets for cancer reversion and predictive, preventive, precision oncology.\\


\end{abstract}

\section{Introduction}

Pediatric high-grade gliomas (pHGGs), such as diffuse midline gliomas (DMGs, or DIPGs in previous nomenclature) and glioblastomas (GBM) are complex adaptive ecosystems and are among the leading causes of cancer-related deaths in Western children\cite{Saratsis2024, jang2025}. They exhibit a median survival of 1–2 years, primarily due to their maladaptive behaviors, including, intratumoral (molecular) heterogeneity, limited treatment penetration across the blood-brain barrier, high rates of recurrence, invasion, and therapy resistance, driven by multiscale regulatory networks \cite{jang2025, wang2025}. Much of the resilience and aggressiveness are attributed to their phenotypic plasticity, i.e., the adoption of a fluid spectrum of stem-like or progenitor-like phenotypic states with state-transitions driving disease progression \cite{Gimple2022,Hanahan2022, Larsson2024,wang2025}. Plasticity acts as the "evolvability engine" and cognitive scaffold shaping cell fate cybernetics, giving rise to emergent, maladaptive behaviors \cite{Uthamacumaran2025}. Despite debated differences in cells of origin, including developmentally arrested progenitor states, complex attractor dynamics from single-cell multiomics reveal a fluid, hybrid identity with a teleonomic bias toward specific terminal fates \cite{jessa2019, jessa2022, uthamacumaranzenil2022, Uthamacumaran2025}.  
However, these cell fate trajectories remain incomplete due to differentiation blockade, preventing attainment of their teleonomic endpoints (cell fate commitments) \cite{jessa2019, jessa2022, uthamacumaranzenil2022, Uthamacumaran2025}.The behavioral patterns emerging from these arrested developmental processes, such as dysregulated cell fate decisions, are complex dynamical systems driven by plasticity networks \cite{uthamacumaranzenil2022}.

Plasticity markers/signatures, or transition genes, embedded within transcriptional regulatory networks enable cancer ecologies to creatively adapt to environmental perturbations, forming resilience through (mal)adaptive behavioral patterns and aberrant signaling dynamics. The absence of safe, and effective targeted therapies, capable of constraining malignant plasticity, underscores the urgent clinical need for causal and predictive biomarker discovery to forecast plasticity markers \cite{Hanahan2022,mehta2024}. These plasticity markers, which denote cell fate transition or bifurcation signatures along pathological attractor dynamics, hold translational potential as therapeutic targets for advancing prevention, precision, and prediction in patient care\cite{Hanahan2022,mehta2024}.

Achieving this requires complex dynamical systems approaches integrated with predictive causal inference algorithms and network science to forecast cell fate cybernetics (i.e., decision making) \cite{uthamacumaranzenil2022}. Complex systems theory, or simply complexity science, is the study of emergent collective behaviors in interconnected dynamical systems, encompassing complex networks and nonlinear dynamics \cite{uthamacumaran2020, uthamacumaran2021}. It frames cancer ecosystems as network processes of multi-scale, system-level interactions and dynamic relationships, where the whole exhibits adaptive behaviors beyond the sum of its parts \cite{uthamacumaran2022alg,Uthamacumaran2025}. In such systems, every agent or sub-part encodes traces of the whole, and the whole is interdependent and interwoven; identity emerges only in relation to other parts, through reciprocal agent–environment interactions as feedback loops. Thence, cancer ecology can only be understood as units of interactions, irreducible to isolated components or agents.

Through this framework, forecasting cell fate decisions and emergent behavioral patterns from plasticity networks parallels weather forecasting in complex, chaotic systems \cite{uthamacumaran2020, uthamacumaran2021}. This  perspective necessitates transdisciplinary frameworks from nonlinear dynamics, such as chaos theory, bifurcations, attractor dynamics, criticality, and network theory, to capture phase transitions and behavioral patterns in cancer networks \cite{uthamacumaranzenil2022}. As such, algorithmic information dynamics (AID) serves as a complexity-informed causal inference framework, to decode and predict the attractor landscapes underlying cancer state-transitions \cite{Zenil_Causal2019,uthamacumaranzenil2022}. AID provides a computational medicine toolkit for reconstructing transition and bifurcation events, mapping attractor dynamics, and inferring plasticity signatures \cite{Zenil_Causal2019}. By network perturbation analysis, algorithmic or graph-complexity methods decode the causal information dynamics of transcriptional processes driving cell-fate trajectories\cite{Zenil_Causal2019}.

In this study, we trace glioma plasticity to epigenetic and developmental (lineage) plasticity by identifying plasticity signatures that steer and regulate cell fate transitions in glioma ecosystems. Rather than following the orchestrated developmental processes and differentiation hierarchies of healthy morphogenesis (pattern formation), pHGGs navigate a decentralized, multilineage attractor landscape that permits adaptive state-transitions and intratumoral heterogeneity, with a potential bias toward NPC/neuronal-like identities (see Supplementary Information).

pHGGs possess well-characterized driver mutations that disrupt neurodevelopmental patterning processes \cite{jessa2019, jessa2022}. Yet, despite advances in AI-driven cancer multiomics, lineage mapping, and data science, the control of differentiation trajectories in pHGGs, such as glioblastoma and DMGs, remains poorly understood \cite{xiang2025, wang2025, gong2025}. While many genetic and epigenetic drivers are known, these are not necessarily plasticity signatures—predictive biomarkers of cell fate transitions \cite{uthamacumaran2022alg, uthamacumaranzenil2022, Uthamacumaran2025}. We need causal inference algorithms to identify critical network biomarkers and drivers that steer cell fate bifurcations.  Decoding these markers and their underlying networks enables forecasting of cell state transitions and identification of key control nodes for phenotypic reprogramming. Addressing the reverse control of stalled developmental programs (differentiation blockade) could pave the way for early detection and safer precision therapies, such as epigenetic or cell fate reprogramming \cite{bizzarri2020, proietti2022}. This resonates with the paradigm of “differentiation therapies,” or cancer reversion therapies, aiming to constrain malignant plasticity and redirect cancer cell fates toward stable attractor states (terminal differentiation or lineage trans-differentiation).

By “attractors,” we refer to long-term behavioral patterns in the symbolic gene expression decision space that steer cell fate trajectories. Studying these requires analysis of “collective intelligence” in tumor ecologies—multiscale molecular-to-cellular networks that govern developmental processes, attractor dynamics, and emergent behaviors driving disease progression. Predicting the regulators of such trajectories is a complex problem requiring dynamic, multiscale modeling. We hypothesize that complex systems theory  can predict and decipher the regulatory dynamics underlying disrupted differentiation hierarchies, niche construction, and transcriptional-level plasticity \cite{Coffey1998, uthamacumaranzenil2022, shin2023}. This knowledge gap underscores the need to examine how fetal brain patterning and self-organization are altered in glioma ecosystems \cite{baig2024holistic, winkler2023cancer}.

To decode latent plasticity regulators across pHGG subtypes, we applied AID—a causal discovery framework quantifying network complexity and dynamic changes under in silico perturbations \cite{Zenil_Causal2019, Zenil_Decomp2018}. AID conceptualizes cancer as a complex adaptive system in which shifts in algorithmic (Kolmogorov) complexity serve as biomarkers of network instability and maladaptive behaviors, including critical phase transitions that drive malignant plasticity \cite{zenil2019, uthamacumaranzenil2022}. By measuring how perturbations alter complexity, AID infers causal contributions of nodes, links, and modules within gene regulatory networks, linking these perturbation responses to attractor dynamics and revealing plasticity signatures.

Unlike traditional statistical or entropy-based approaches, AID is model-agnostic and integrates measures such as the Block Decomposition Method (BDM) and lossless compression algorithms (e.g., LZW) to detect deep algorithmic regularities and causal dependencies \cite{Zenil_Causal2019, Zenil_Decomp2018}. Integrated with Non-negative Matrix Factorization (NNMF) for dimensionality reduction, AID identified plasticity markers that forecast glioma cell state transitions, pinpointing regulatory structures that steer differentiation arrest, lineage bias, and transdifferentiation potential \cite{uthamacumaranzenil2022}. These perturbation-based attractor dynamics reconstruction reveal critical tipping points and bifurcation signatures (i.e., plasticity regulators) in the transcriptomic decision space, offering network-informed biomarkers for precision medicine strategies aimed at reprogramming cancer cell fate trajectories.

\section{Methods}

\subsection*{Bulk RNA-seq cohorts and preprocessing}
We analysed bulk RNA-seq data from the public series \textbf{GSE162976} (pediatric high-grade glioma; DIPG and GBM) \cite{wang2021}. The dataset comprised human neural stem cells (hNSC, denoted here as “normal” controls) and patient-derived pediatric glioma cell lines (DIPG and GBM subtypes), all sequenced on the Illumina NextSeq 500 platform and aligned to hg38.  

All conditions, except for the DIPG007 sample, contained \textbf{three independent biological replicates}, including hNSC controls (normal), DIPG cell lines (DIPGXIII, DIPGIV, SF8628, SF7761, DIPG1114, and DIPG007 with two replicates), and GBM cell lines (KNS42, SF9427, SF9402). For our primary contrasts, we used \textbf{3 hNSC replicates} as the normal (control) baseline, \textbf{17 DIPG samples} (including one DIPG tissue specimen within the DIPGXIII set), and \textbf{9 GBM samples}. In our analyses, the "Combined Glioma” refers to the merged DIPG+GBM group (26 glioma samples) compared against hNSC. Processed count matrices from GEO were imported, and merged into a normalized expression matrix for downstream DESeq2 analysis.

\subsection*{Differential expression analysis (DESeq2)}
Differentially expressed genes (DEGs) were computed with \texttt{DESeq2} (R, v$\geq$1.38) \cite{love2014deseq2} comparing DIPG, GBM, or combined glioma samples against human neural stem cells (hNSC) as healthy controls, using contrasts (i) DIPG vs.\ hNSC , (ii) GBM vs.\ hNSC, and (iii) combined glioma (DIPG+GBM) vs.\ hNSC. Size factors were estimated by the median-of-ratios method; dispersions were fitted using the default parametric trend; Wald tests were applied to log fold-changes. Multiple testing was controlled by Benjamini–Hochberg; genes were retained as DEGs if adjusted $p$-value ($\mathit{padj}$) $<0.05$ and $|\log_{2}\mathrm{FC}|\geq 2$. Normalised counts produced by DESeq2 were exported for network construction and downstream analyses. The resulting filtered DEGs were then used to construct adjacency matrices from the normalized count expression matrices. Specifically, for each set of filtered genes, a Spearman correlation coefficient and a Partial Information Decomposition (PID) score were computed between the samples (columns) within the normalized counts matrix.

\subsection{Algorithmic Information Dynamics, Graph Construction, and Network Metrics}

Algorithmic Information Dynamics (AID), grounded in Algorithmic (Kolmogorov–Chaitin) complexity, provides a distribution-free, scale-invariant framework for linking network topology to intrinsic computational structure and causal content. Unlike purely statistical metrics, AID captures nonlinear, emergent, and goal-directed dynamics without reliance on probability distributions, enabling a universal assessment of functional information in complex systems \cite{zenil2019}. Kolmogorov complexity estimates the shortest possible description length of a network, reflecting its intrinsic computational structure, while revealing the interplay between graph network architecture, function, and adaptive behaviour.

For each contrast, we constructed sample–sample and gene–gene adjacency matrices from DESeq2-normalised counts using Spearman correlation; in supplementary analyses, we incorporated Partial Information Decomposition (PID) to capture redundancy/synergy structures. Two types of adjacency matrices—Spearman- and PID-derived—were analysed within the algorithmic information theory framework.

We applied the Block Decomposition Method (BDM) using the \texttt{pyBDM} package to compute algorithmic complexity on binarised adjacency matrices (threshold = 0.5), quantifying information content by decomposing networks into smaller substructures and approximating their Kolmogorov complexity. Shannon entropy was computed to measure uncertainty in network connectivity, while Lempel–Ziv–Welch (LZW) compression (\texttt{zlib} implementation) served as an alternative proxy for global complexity, capturing redundancy and regularity.

AID-based perturbation analysis was performed by systematically removing nodes/edges and recalculating BDM to estimate algorithmic perturbation changes (BDM shifts), thereby ranking genes and links by causal contribution to global network (attractor) stability. This method identifies topological features—analogous to centrality measures—that drive robustness, resilience, and attractor dynamics. Such attractor states represent regions in gene-expression state-space that govern stable or unstable system behaviours unless perturbed. Genes with consistently high perturbation scores (large $\Delta$BDM under removal) and supportive network centrality were designated as \emph{plasticity signatures}, representing potential biomarkers or therapeutic targets in glioma cell-state transitions and reprogramming.

Standard network centrality metrics—including degree, betweenness, closeness, eigenvector centrality, and hub score—were also computed to evaluate influence and connectivity: closeness measures efficiency of information flow, betweenness identifies communication bottlenecks, eigenvector centrality highlights nodes linked to other influential nodes, and hub score pinpoints major interaction hubs \cite{evans2022linking}. Network topology and dynamics were visualised using \texttt{NetworkX} to examine structural organisation, hierarchical patterns, and functional modules within the inferred transcriptional regulatory networks. In addition to network (adjacency matrix) perturbation analysis, complexity measures such as BDM and lossless compression (LZW) were computed on the network centrality metrics themselves, providing complementary patterns and features steering graph information flow. This  allowed us to identify key information pathways and structural bottlenecks, highlighting critical nodes whose connectivity patterns contribute disproportionately to the network’s plasticity, and hence, capacity for signal propagation and cell fate control.

\subsection{ Single-cell scRNA-Seq Datasets}

This study utilized publicly available scRNA-seq datasets from two foundational studies characterizing the transcriptional heterogeneity of HGGs. We refer to samples from Neftel et al. (2019) as glioblastoma, and those from Filbin et al. (2018) as DIPG, consistent with the terminology used in their respective publications \cite{neftel2019integrative, filbin2018developmental}. The pediatric glioblastoma dataset (GSE131928) includes the TPM values (transcript-per-million reads) of 8 patients with IDH-wildtype glioblastoma, profiled using Smart-seq2 (7,930 cells) to capture gene expression states across diverse tumor subpopulations. The DIPG dataset (GSE102130) comprises 4,058 single cells, primarily derived from six H3K27M-mutant diffuse midline gliomas, and was also sequenced using the Smart-seq2 platform. Processed data matrices were log-transformed prior to analyses. 

\subsection{Single-cell Analysis}

All of the above-mentioned methods were repeated for single-cell datasets to compare with the bulk analysis. Two single-cell gene expression matrices (cells by genes) were utilized: one for pediatric IDH wild-type glioblastoma (IDHWT) and the other for K27M DIPG. Differentially expressed genes (DEGs) between these two datasets were identified using a two-sample t-test between groups, and p-values were adjusted for multiple comparisons using the Benjamini-Hochberg method (False Discovery Rate, FDR). Genes with adjusted p-values (\textit{padj}) $<$ 0.05 and \(|\log_2\text{FoldChange}| \geq 1\) were selected as DEGs. From these, the top 100 DEGs based on absolute \(\log_2\)FoldChange values were selected for downstream analyses such as BDM perturbation analysis. These results consisted of the 'combined glioma' analysis for the single-cell datasets. For individual group analysis, therefore, GBM and DIPG on their own, non-negative matrix factorization (NNMF) was applied to identify transcriptional modules within each dataset, and the top 50 genes per module were selected for additional downstream analyzes.

NNMF decomposes the gene expression matrix into non-negative "metagenes" expression patterns, i.e., hierarchically clustering genes into modules based on shared expression profiles by use of loadings (weights) assigned by variance explained. Top NNMF modules were selected based on these weights, variance, and sparsity, capturing dominant transcriptomic patterns across the glioma subtypes. 

Adjacency matrices were constructed using pairwise Spearman correlations of the expression data, followed by binarization at thresholds of 0.1 and 0.5. Network-level analyses were conducted using the NetworkX library to calculate centrality measures, including closeness, betweenness, and eigenvector centrality, as topological signatures predicting cell fate transition dynamics. These centrality-based network biomarkers identify critical regulatory hubs and drivers of network dynamics, serving as plasticity regulators of cell fate cybernetics.

Perturbation analysis of single genes and gene-gene links was performed using the BDM package, which quantifies the complexity of binary matrices. This included analyzing the effects of node and link perturbations on network complexity. 
\begin{figure}[H]
    \centering
    \includegraphics[width=0.6\textwidth]{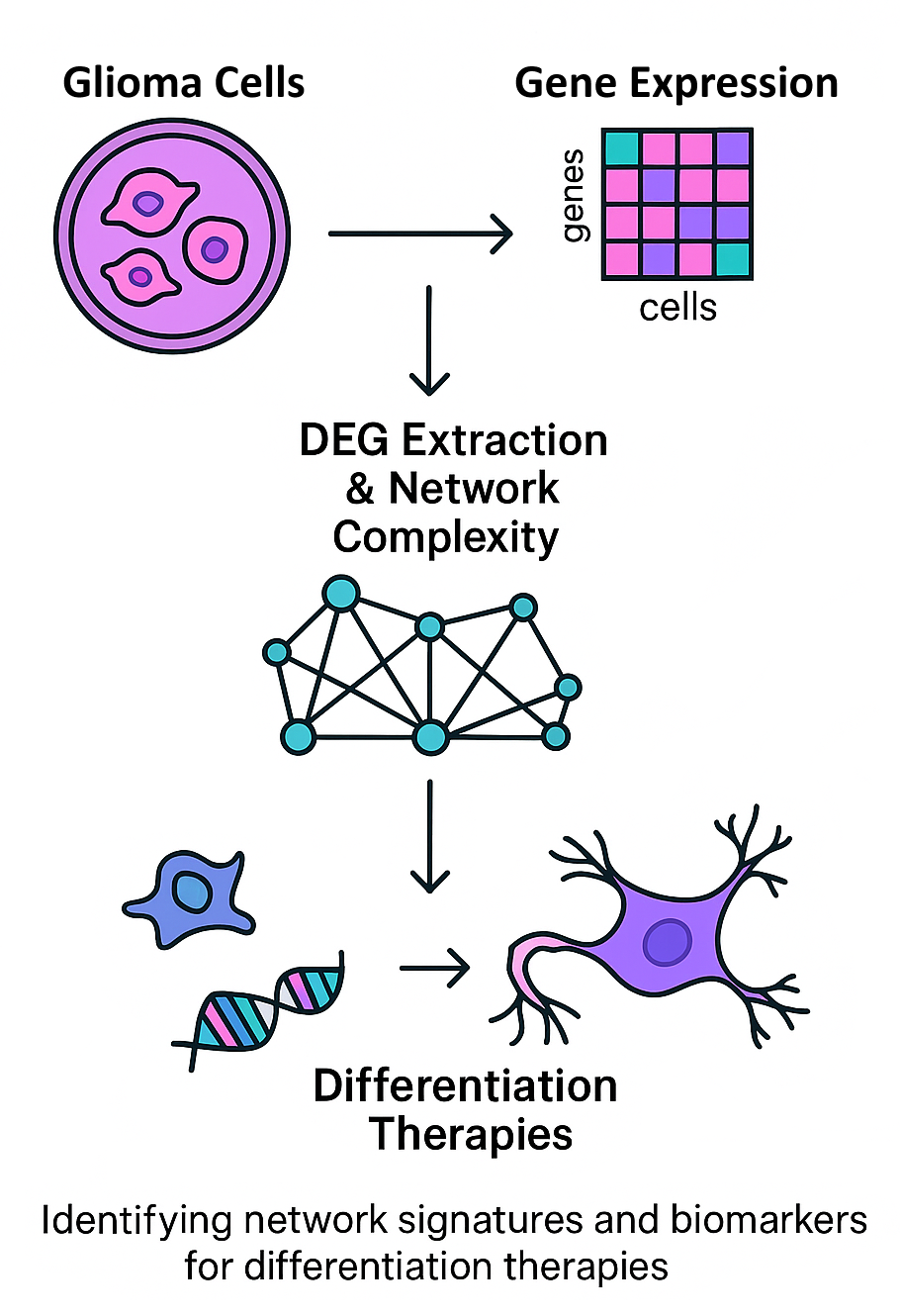}
    \caption{\textbf{Workflow schematic for biomarker discovery using algorithmic graph complexity and network signatures.} 
    This diagram illustrates the stepwise pipeline from glioma cell transcriptomics to decoding differentiation gene expression patterns (DEG), and BDM-based network complexity analysis, enabling identification of cellular reprogramming targets  in precision medicine. The goal is to drive plastic malignant states toward stable neuron-like cellular identities with translational oncology potential.}
\label{fig:glioma_complexity_workflow}
\end{figure}

\section{Results}

\subsection{Neurodevelopmental Plasticity and Bioelectric Patterning Signatures Define Glioma Transcriptional Networks}

\begin{figure}[H]
\centering
\includegraphics[width=0.95\textwidth]{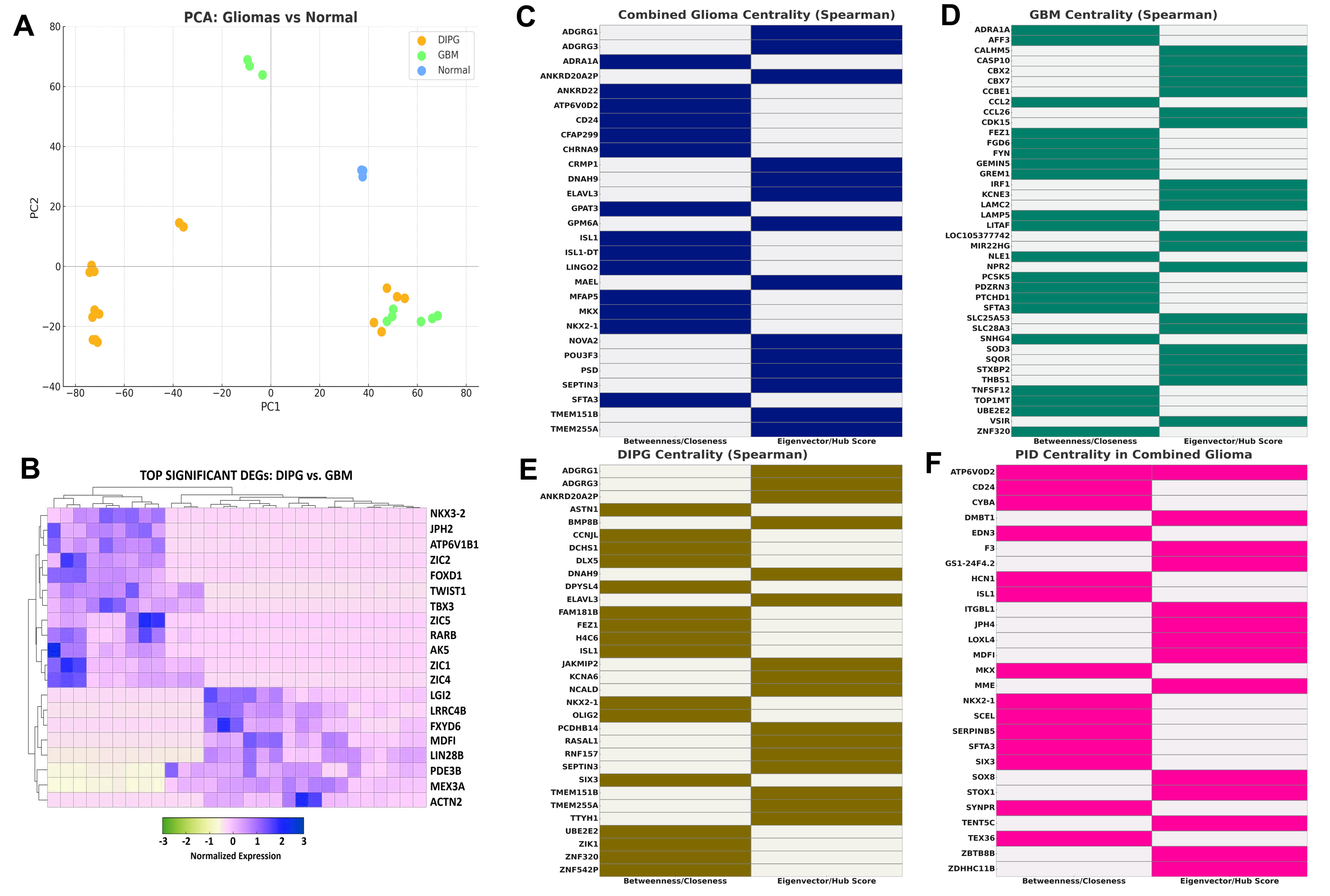}
\caption{\textbf{A.} PCA plot showing clustering of bulk RNA-seq data. Normal healthy brain samples (pink) form a compact cluster; DIPG (teal) and pGBM (violet) display wide dispersion, indicating cellular heterogeneity and multilineage bifurcations.
\textbf{B.} Heatmap of the top 20 DEGs between DIPG and GBM.
\textbf{C–E.} Top-ranked genes in Spearman correlation networks across multilineage bifurcations in combined gliomas (C), GBM (D), and DIPG (E), highlighting developmental regulators and bioelectric/neuroimmune modulators.
\textbf{F.} PID network analysis in combined gliomas, identifying unique mediators not captured by correlation networks.}
\label{fig:pca_deg_merged}
\end{figure}

The principal component analysis (PCA) of bulk RNA-seq data revealed distinct clustering of normal healthy brain samples (pink) in a compact region, while DIPG (teal) and pGBM (violet) samples displayed a wide dispersion in PCA space (Figure \ref{fig:pca_deg_merged}A). This dispersion reflects pronounced cellular heterogeneity, consistent with a fluid continuum of cellular states and multilineage bifurcations within glioma ecosystems. Differential expression analysis between DIPG and GBM identified the top 20 differentially expressed genes (DEGs) (Figure \ref{fig:pca_deg_merged}B). Network centrality measures from glioma correlation analyses highlighted key developmental regulators—POU3F3, NKX2-1, ISL1—and bioelectric/neuroimmune modulators—ADRA1A, CALHM5, LAMC2—as central nodes across multilineage bifurcation contexts (Figures \ref{fig:pca_deg_merged}C–E). Partial Information Decomposition (PID) networks further revealed unique mediators absent from correlation-based graphs, including SIX3, SOX8, JPH4, and ZBTB8B (Figure \ref{fig:pca_deg_merged}F).

Table \ref{tab:top_deg_gliomas} summarizes the top 30 upregulated and downregulated DEGs distinguishing gliomas from normal human neural stem cell (hNSC) transcriptomes. Upregulated DEGs were enriched in epithelial-to-mesenchymal transition (EMT) processes, bioelectric signaling, and embryonic development. Many were developmental transcription factors (NKX2-1, ISL1, SIX3) and patterning regulators, along with neuronal/synaptic genes (SYNPR, HCN1, CHRNA9, MYO3A) that serve as bioelectric regulators. Downregulated DEGs included bioelectric patterning markers such as voltage-gated and ligand-gated ion channels (OLIG2, GRIK3, CACNG5, KCNJ16, GABRA5), solute transporters, and inhibitory neurotransmitter receptors. Other signatures—GRIK3, GABRA5, SLC24A3, KCNA6, SHISA6, KCNK10, FOXR2, SLC8A3, CADM2, CSMD1—reflect disruption of the bioelectric code and altered neuron–glioma interactions, potentially steering stalled differentiation programs and phenotypic switching (multilineage plasticity).

Some of the most striking fold changes ($>$27-fold) occurred in OLIG2 and GRIK3, highlighting a profound shift in neurodevelopmental landscape and lineage hierarchy along a ventral–dorsal patterning gradient. These transcriptional  signatures confirm that glioma ecosystems operate within altered morphogenetic processes, disrupted bioelectric patterning, and convergent multilineage plasticity that shape tumor phenotypic heterogeneity. However, the predominance of neuronal and synaptic markers among the upregulated genes suggests a lineage or cell fate bias toward neuronal-like differentiation programs, indicating that glioma cells may be navigating toward or adopting neurodevelopmental and neuronal-like identities as a result of differentiation arrest.

\begin{table}[H]
    \centering
    \caption{Top 30 upregulated and downregulated DEGs distinguishing gliomas (GBM and DIPG) from normal hNSC bulk RNA transcriptomes.}
    \label{tab:top_deg_gliomas}
    \begin{tabular}{lcc|lcc}
        \multicolumn{3}{|c|}{\textbf{Top 30 Upregulated DEGs}} & \multicolumn{3}{c|}{\textbf{Top 30 Downregulated DEGs}} \\
        \hline
        \textbf{Gene} & \textbf{log2FoldChange} & \textbf{padj} & \textbf{Gene} & \textbf{log2FoldChange} & \textbf{padj} \\
        \hline
        LOC105376188 & 10.7551 & 4.16E-45 & OLIG2 & -27.3507 & 1.39E-21 \\
        SFTA3 & 10.6526 & 5.92E-07 & GRIK3 & -27.2121 & 6.30E-20 \\
        LOC102724934 & 10.4227 & 1.48E-19 & CACNG5 & -27.0883 & 2.70E-24 \\
        SYNPR & 9.9399 & 0.0016 & ASTN1 & -27.0706 & 2.32E-24 \\
        SIX3 & 9.7709 & 3.89E-11 & DCX & -26.7213 & 1.15E-12 \\
        MKX-AS1 & 9.0274 & 0.1320 & AMER2 & -26.4478 & 1.05E-15 \\
        ISL1 & 8.9041 & 3.53E-05 & SOX8 & -26.3295 & 8.89E-24 \\
        MKX & 8.8226 & 6.96E-06 & TFAP2C & -25.9380 & 6.35E-24 \\
        EDN3 & 8.7319 & 4.83E-06 & CXCL6 & -25.2102 & 6.11E-18 \\
        NKX2-1 & 8.6624 & 0.0003 & DAB1 & -25.1401 & 4.43E-22 \\
        HCN1 & 8.6218 & 1.39E-09 & CADM2 & -25.0891 & 6.35E-24 \\
        ATP6V0D2 & 8.5907 & 9.21E-12 & KCNJ16 & -25.0446 & 7.83E-18 \\
        TEX36 & 8.2290 & 2.39E-11 & MXRA5 & -24.9576 & 1.81E-17 \\
        SIX3-AS1 & 8.0981 & 6.38E-11 & GABRA5 & -24.7729 & 3.29E-17 \\
        SCEL & 8.0542 & 1.38E-09 & SLC8A3 & -24.7149 & 1.92E-22 \\
        NKX2-1-AS1 & 7.4257 & 0.0068 & LINC00643 & -24.6687 & 4.16E-14 \\
        SERPINB5 & 7.3041 & 0.0012 & LOC107985505 & -24.5553 & 1.52E-16 \\
        GUCA1C & 7.1677 & 0.0736 & TOX3 & -24.4346 & 1.52E-12 \\
        CD24 & 6.9694 & 0.0004 & ELAVL4 & -24.4017 & 8.00E-16 \\
        CYBA & 6.9213 & 1.91E-10 & TMPRSS15 & -24.4008 & 6.50E-12 \\
        ADRA1A & 6.9053 & 0.0025 & ZDHHC22 & -24.3467 & 6.11E-18 \\
        MYO3A & 6.8896 & 0.0077 & TSPYL5 & -24.3157 & 6.30E-19 \\
        ISL1-DT & 6.7336 & 0.0129 & IGSF11 & -24.2593 & 1.27E-18 \\
        CYSLTR2 & 6.6542 & 1.70E-06 & TNFSF15 & -24.1712 & 1.59E-15 \\
        LINC02849 & 6.5251 & 4.34E-19 & ADARB2 & -24.0821 & 4.94E-12 \\
        ANKRD22 & 6.4570 & 0.0002 & FOXB1 & -23.9952 & 7.18E-16 \\
        MFAP5 & 6.4027 & 0.0011 & MYH14 & -23.9355 & 1.08E-19 \\
        CHRNA9 & 6.2258 & 0.0010 & EN1 & -23.8609 & 7.76E-15 \\
        
    \end{tabular}
\end{table}

\subsection{Network Centrality and Complexity Metrics Identify Developmental and Bioelectric Drivers of Glioma Plasticity}

To complement Spearman-based co-expression networks, we evaluated four complementary centrality measures, such as betweenness, closeness, eigenvector, and hub score, across glioblastoma (GBM), diffuse intrinsic pontine glioma (DIPG), and combined glioma networks (Figure~\ref{fig:pca_deg_merged}). These analyses highlight candidate regulators of glioma subtype–specific plasticity dynamics, lineage transitions, and differentiation pathways.

\textbf{Combined Glioma Network (Panel A):} High-ranking nodes included \textit{POU3F3}, \textit{NKX2-1}, \textit{ISL1}, and \textit{ADGRG1}, all previously implicated in neurodevelopmental signaling and cortical cell migration, reinforcing their role in differentiation and plasticity. \textit{SFTA3}, functionally linked to \textit{NKX2-1}, was also prominent, with known roles in wound healing and neuroendocrine morphogenesis.

\textbf{GBM-Specific Network (Panel B):} Centralities emphasized a blend of bioelectric and immune-related regulators. \textit{ADRA1A} (adrenergic receptor) and \textit{CALHM5} (calcium homeostasis modulator) point to electrochemical signaling functions, while \textit{GREM1} and \textit{LAMC2} participate in BMP signaling and extracellular matrix remodeling, potentially influencing invasiveness. Immune-linked hubs such as \textit{VSIR}, \textit{CCL2}, and \textit{CASP10} highlight interplay between immune suppression and tumor growth. GBM displayed the highest complexity measures (BDM = 869.027, Entropy = 0.0748, LZW = 94{,}947{,}884), indicating substantial transcriptional irregularity and network diversity.

\textbf{DIPG-Specific Network (Panel C):} This network recaptured shared developmental regulators (\textit{NKX2-1}, \textit{UBE2E2}, \textit{ISL1}) but uniquely featured forebrain development, neurulation, and neural crest differentiation factors such as \textit{DLX5}, \textit{SIX3}, \textit{OLIG2}, and \textit{ZNF320}. Neurodevelopmental transcription factors \textit{ELAVL3} and \textit{ANKRD20A2P} were also enriched, reflecting the epigenetic reprogramming landscape of pediatric gliomas. g:Profiler enrichment identified nervous system development ($p.adj = 1.165 \times 10^{-3}$), with \textit{DLX5}, \textit{OLIG2}, and \textit{ISL1} predictive of neural crest differentiation ($p.adj = 4.211 \times 10^{-2}$). Additional significantly enriched TFs included BTEB3, MED8, GKLF, WT1, and TFII-I. DIPG had the lowest complexity measures (BDM = 703.17, Entropy = 0.0629, LZW = 79{,}130{,}418), suggesting more constrained developmental programs and lineage-transition dynamics.

\textbf{Information-Theoretic PID Network (Panel D):} To capture causal dependencies, we constructed a Partial Information Decomposition (PID) network for combined gliomas. PID decomposes shared information into unique, redundant, and synergistic components, revealing causal and directional information flow in multivariate systems. This analysis identified developmental and metabolic regulators such as \textit{SIX3}, \textit{SOX8}, and \textit{TENT5C}, linked to embryonic patterning, axon guidance, and morphogenetic compartmentalization. Additional bioelectric and signaling hubs included \textit{ATP6V0D2}, \textit{DMBT1}, \textit{JPH4}, and \textit{STOX1}. Several genes detected here---\textit{ZDHHC11B}, \textit{ZBTB8B}, \textit{CYBA}---were absent from Spearman networks, underscoring PID’s ability to uncover hidden, nonlinear regulatory nodes.

\textbf{Additional Cross-Network Targets:} Beyond the main panels, we identified immunotherapeutic and neurotransmitter-linked targets of potential therapeutic relevance. These included immune escape mediators \textit{CCL26}, \textit{VSIR}, \textit{IRF1}, and bioelectric/neurotransmitter circuit components \textit{ADRA1A}, \textit{KCNA6}, \textit{FYN}, and GRIN-like channels, implicating membrane potential and neurotransmission in network modularity and glioma vulnerability.

Collectively, these centrality and complexity metrics reveal that GBM harbors the most structurally and algorithmically complex transcriptional networks, consistent with heightened aggressivity, while DIPG maintains more developmentally constrained, lineage-biased dynamics. The identified hubs converge on neurodevelopmental patterning and bioelectric regulation as core drivers of glioma plasticity.

\subsection{BDM Perturbation and PID Network Analysis Reveal Neurodevelopmental, Immune, and Metabolic Drivers of Glioma Plasticity}

\begin{figure}[H]
\centering
\includegraphics[width=1.0\textwidth]{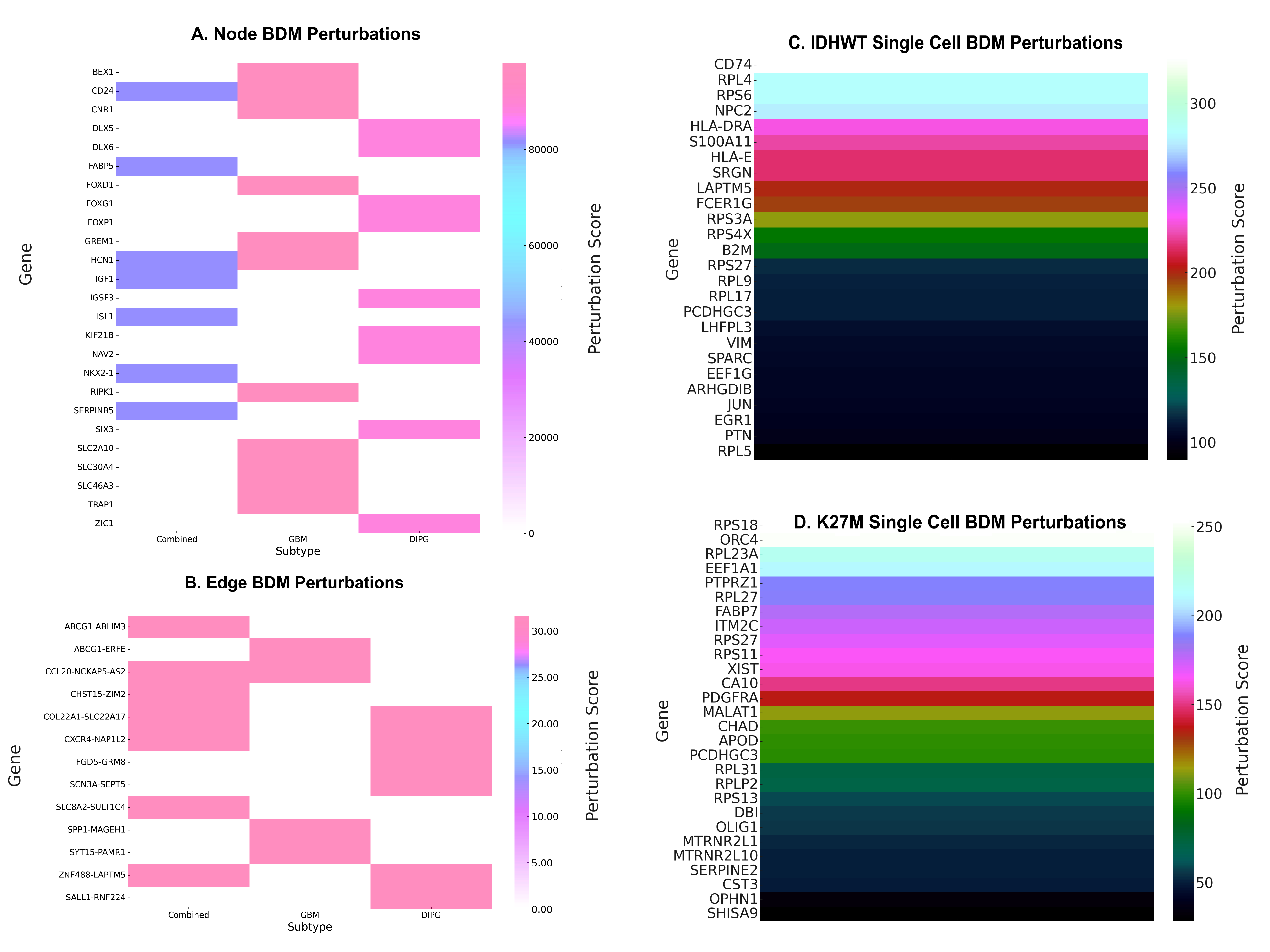}
\caption{\textbf{A.} Node BDM perturbations across glioma subtypes highlight neurodevelopmental, immune, and metabolic regulators. \textbf{B.} Edge BDM perturbations across glioma subtypes reveal rewired synaptic signaling, ECM remodeling, and immune modulation. Color bars indicate perturbation strength.}
\label{fig:bdm_nodes}
\end{figure}

We investigated the topological impact of differentially expressed genes (DEGs) using BDM perturbation scores derived from Partial Information Decomposition (PID) networks across glioma subtypes (Combined, GBM, and DIPG). These scores quantify the causal structures and sensitivity of each node or edge to expression variation, enabling a systems-level understanding of network fragility and emergent control points. Figure~\ref{fig:bdm_nodes} presents the most significantly perturbed nodes (individual genes) and edges (gene pairs) across subtypes. Gene functions were annotated using the GeneCard Database.

\paragraph{Node BDM Perturbations (Figure~\ref{fig:bdm_nodes}A).}
In the combined glioma PID network, \textit{ISL1} and \textit{NKX2-1} (both 81{,}508 bits) emerged as top BDM-shifted genes. These neurodevelopmental transcription factors are implicated in neuronal lineage specification and cell fate commitment. Additional highly perturbed genes included \textit{IGF1} and \textit{FABP5}, regulators of growth signaling and lipid metabolism, as well as immune modulators \textit{SERPINB5} and \textit{CD24}, both linked to tumor progression and immune evasion. \textit{NKX2-1} (thyroid transcription factor-1, TTF-1), with roles in neurodevelopment, was observed as a combined glioma signature.

In the GBM network, the most perturbed genes included \textit{CD24}, \textit{RIPK1}, and \textit{TRAP1}, all involved in inflammatory cytokine signaling, mitochondrial stress regulation, and immune modulation. \textit{FOXD1} and \textit{BEX1}, both morphogenesis regulators, were uniquely perturbed in GBM. \textit{BEX1}, the top GBM node hit, encodes a signaling adaptor in the p75NTR/NGFR pathway, linking neurotrophic signaling to cell cycle regulation and neuronal differentiation, implicating it in glioma plasticity. \textit{CNR1} (cannabinoid receptor) further highlighted a neuro–immune axis in GBM plasticity. \textit{GREM1}, an antagonist of BMP signaling involved in developmental patterning, also appeared as a high BDM signature in GBM.

In DIPG, perturbations centered on neurodevelopmental regulators such as \textit{DLX5}, \textit{DLX6}, \textit{ZIC1}, \textit{SIX3}, and \textit{FOXG1}, which orchestrate early brain patterning and differentiation. These findings reinforce the view that DIPG tumors remain in a developmentally plastic, partially differentiated state. \textit{FOXP1} and \textit{IGSF3} were also perturbed, suggesting chromatin remodeling and immune–neurodevelopmental interplay. FOX clusters appeared subtype-specific: \textit{FOXD1} was among the highest single-node BDM shifts in IDHWT GBM, while \textit{FOXG1} and \textit{FOXP1} were observed in DIPG.

Axon guidance genes \textit{NAV2} and \textit{KIF21B} further indicated misregulated neuronal wiring within the tumor microenvironment. \textit{HCN1}, a hyperpolarization-activated cyclic nucleotide-gated ion channel, was commonly perturbed in both GBM and Combined networks, implicating bioelectric dysregulation.

Additional GBM-specific metabolic perturbations included \textit{SLC2A10}, \textit{SLC46A3}, and \textit{SLC30A4}, involved in glucose and folate transport and metal ion regulation. Together with the neurodevelopmental and immune signatures, these disruptions suggest that BDM perturbation can pinpoint subtype-specific vulnerabilities. Notably, BDM perturbations in PID networks recapitulated plasticity signatures identified in Spearman correlation-based DEG networks, including immune modulators, neurodevelopmental factors, and metabolic regulators. This convergence underscores BDM’s utility for detecting structurally and functionally critical nodes and links in causal biomarker discovery and systems medicine strategies for glioma plasticity.

\paragraph{Edge (Link) BDM Perturbations (Figure~\ref{fig:bdm_nodes}B).}
In the Combined glioma network, highly perturbed edges included \textit{CHST15}–\textit{ZIM2} and \textit{CXCR4}–\textit{NAP1L2}. \textit{CHST15} modifies chondroitin sulfates in the extracellular matrix, reflecting niche construction and structural remodeling of the tumor–immune microenvironment. \textit{CXCR4} mediates immune cell trafficking and neuron–glia interactions, reinforcing immune plasticity themes.

In GBM, top edge perturbations included \textit{SPP1}–\textit{MAGEH1} and \textit{CCL20}–\textit{NCKAP5-AS2}, reflecting inflammatory signaling, epigenetic modulation, and proneural-to-mesenchymal transitions. \textit{SYT15}–\textit{PAMR1} highlighted neuronal identity and synaptic pathways active in GBM’s plastic microenvironment.

In DIPG, highly perturbed pairs included \textit{GRM8}–\textit{FGD5}, \textit{SCN3A}–\textit{SEPT5}, and \textit{ZNF488}–\textit{LAPTM5}. \textit{GRM8} encodes a metabotropic glutamate receptor essential for synaptic plasticity, while \textit{SCN3A} encodes a voltage-gated sodium channel critical for neuronal excitability. \textit{FGD5} and \textit{COL22A1} support vascular remodeling and neural tube development, implicating endothelial and matrix pathways. \textit{SALL1}, appearing as a perturbed edge with \textit{RNF224}, points to chromatin remodeling during early lineage transitions.

Bioelectric and transport-related edges included \textit{SLC8A2}–\textit{SULT1C4} (calcium ion exchanger and sulfotransferase) and \textit{NAV2}–\textit{GRPR} (axonal extension and neuropeptide signaling). These interactions converge on neuronal circuitry mechanisms hijacked in tumorigenesis. Thus, edge-level BDM perturbation complements node-level findings, offering finer resolution of communication bottlenecks and regulatory rewiring in glioma subtypes, and revealing putative targets for cell fate reprogramming and subtype-specific targeted therapy discovery.

\subsection{Single-Cell BDM Perturbation Analysis Reveals Immunomodulatory and Plasticity Hubs in Glioblastoma and DIPG}

As shown in Figure~\ref{fig:bdm_nodes}C, \textit{CD74} exhibited the highest perturbation score (327.02) in IDHWT glioblastoma single-cell BDM signatures. \textit{CD74} is a key immune modulator involved in MHC class II signaling and antigen presentation, suggesting its role as a complexity-based (BDM) network hub and a potential immunotherapy target through disruption of tumor–immune interactions.

Other notable IDHWT-specific perturbation signatures included \textit{S100A11}, \textit{JUN}, \textit{EGR1}, and \textit{ARHGDIB}—genes linked to calcium signaling, stress response, early growth regulation, and immune modulation. Their elevated BDM scores implicate them in niche construction within an aggressive, plastic, and inflammatory microenvironment.

Figure~\ref{fig:bdm_nodes}D shows that in the K27M single-cell PID network, well-established glioma plasticity markers—\textit{PTPRZ1}, \textit{FABP7}, \textit{PDGFRA}, and \textit{OLIG1}—rank among the highest single-node BDM perturbation scores. These genes are associated with gliomagenesis, stemness, and neural lineage specification~\cite{Uthamacumaran2025}. Additionally, mitochondrial genes \textit{MTRNR2L1} and \textit{MTRNR2L10} appeared prominently, potentially indicating oxidative stress adaptation, apoptotic protection, or metabolic signaling via humanin-like peptides.

The heatmap color scale in Figures~\ref{fig:bdm_nodes}C--D reflects perturbation intensity, with lighter (whitish) tones denoting higher BDM perturbation scores and darker hues representing lower scores.

Therefore, \noindent
Panels C and D contrast single-cell BDM perturbation scores for IDHWT and K27M glioma subtypes, respectively. IDHWT (Panel C) shows enrichment of immune and ribosomal markers (e.g., \textit{CD74}, \textit{HLA-DRA}, \textit{RPS6}, \textit{FCER1G}), suggesting immune infiltration and translational regulation. In contrast, K27M (Panel D) reveals perturbations in mitochondrial non-coding RNAs (\textit{MTRNR2L10}, \textit{MTRNR2L1}), neural/glial lineage genes (\textit{OLIG1}, \textit{PDGFRA}), and long non-coding RNAs (\textit{XIST}, \textit{MALAT1}), reflecting disrupted developmental trajectories and lineage-specific plasticity markers. These subtype-specific BDM signatures highlight distinct cell identity and functional state instability across pediatric glioma subtypes. Furthermore, post-translational modifications of deciphered genes' regulatory proteins may also serve as emergent biomarkers of glioma progression~\cite{Pienkowski2023}, paving targets for liquid biopsies, longitudinal monitoring, and real-time (early) detection.

\section*{Discussion}

\subsection{Causal Network Features Decoded Plasticity Signatures Steering Cell Fate Transitions in Glioma Ecologies}

Driven by their malignant plasticity, pHGGs exhibit invasive phenotypes, therapy resistance, and a lack of predictive markers for early recurrence detection, and targeted therapies, contributing to significant unmet patient needs. This underlines the notion that malignant tumors like pHGGs are complex adaptive ecosystems and disorders of cell fate decisions, requiring principles from cellular cybernetics and reverse-engineering control theory to decode and therapeutically reprogram their malignant traits and maladaptive patterns of behavior\cite{uthamacumaranzenil2022, uthamacumaran2022alg,Uthamacumaran2025}. Our AID framework and its identified network biomarkers offer a causal inference approach to decode therapeutic vulnerabilities, by identifying  plasticity regulators, which steer the cancer network (cellular states) toward aggressivity, and hence, reversibly towards reprogrammability and stability \cite{uthamacumaranzenil2022, Uthamacumaran2025}. As such, these plasticity regulators serve as predictive signatures of tumor progression and targets for precision medicine, such as precision gene editing or 'differentiation therapy'—defined here as the phenotypic  or epigenetic reprogramming of malignant cellular identities stuck in a fluid spectrum of adaptive states (i.e., developmental blockade) toward stable, benign-like attractor states. Plasticity, as the cognitive scaffold underlying cellular heterogeneity and unstable fate dynamics, makes differentiation therapy a means of constraining malignant plasticity, and reintegrating the fragmented systemic communication between cells and their environment, i.e., re-steering maladaptive cybernetics toward physiological coherence, thereby serving as a form of ecosystems engineering. 

In evidence, our graph complexity methods forecasted predictive regulators of glioma plasticity and decoded causal communication patterns reflecting complex signaling networks and niche construction signals within glioma ecologies. Many of the plasticity signatures we decoded were enriched in neurulation-associated morphogenetic signals and transcription factors—including WNT, BMP, TGF-β, DLX, FOX and HOX genes—highlighting pHGGs as complex cybernetic systems and disorders of developmental cell fate decisions. These pathways mediate mesoderm–ectoderm cross-talk during neural tube and neural crest formation, underscoring their reactivation in glioma plasticity and cell fate reprogramming.

Using NNMF clustering and BDM-based information dynamics, we captured causal structures and perturbation responses underlying cell fate transitions. Spearman correlation and PID further enriched network inference signatures, with PID resolving directional, redundant, and synergistic flows of cellular communication—ideal for identifying tissue- and cancer-specific regulatory biomarkers, while correlation served as a simpler baseline of transcriptional patterns. Prior to in silico node or link perturbations, the combined glioma DEG network exhibited high algorithmic complexity (BDM: 883.95 bits) and low entropy (0.0719 bits), indicating structured, non-random gene interactions. Its large LZW size (~66.7M) suggests lossless compressibility. Perturbation analysis discovered regulatory and plasticity signatures that map and compress the network’s information dynamics, as predictive features of glioma cell fate decisions. 

As shown in Figure~\ref{fig:pca_deg_merged}, genes identified through centrality measures included key neurodevelopmental and differentiation markers. Gene set enrichment via g:Profiler revealed the top biological processes—cell differentiation, cellular developmental process, and anatomical structure development (\( p_{\mathrm{adj}} < 0.01 \)). Several genes were involved in mechanotransduction and remodeling of the extracellular matrix (ECM) and cytoskeleton within the tumor-immune microenvironment, including THBS1 (focal adhesion dynamics and TGF-β signaling), LAMC2 (cell adhesion and tissue organization), and CCBE1 (lymphangiogenic ECM remodeling). Others, such as CRMP1 and GREM1, may modulate collective behaviors through axon guidance and BMP antagonism, respectively. FOXD1, TWIST1, and NKX3-2—implicated in epithelial-to-mesenchymal transition (EMT), morphogenesis, and spatial brain patterning—reflect glioma’s underlying lineage and fate plasticity. Additional markers, including ATP6V1B1 and FXYD6, suggest altered bioelectric code and electrophysiological states during malignancy. These expression patterns underscore cellular decision-making in patterning processes, and developmental divergence between GBM and DIPG, particularly along the ventral-dorsal axis. Enrichment of genes such as WNT7A, CAMK2, GRIK1/3, FOXA1, FOXP1, NRXN3, KCNK4/10, PAX6, and RARB, among the top features, further supports a core regulatory program involving synaptic signaling, cell fate commitment, embryonic morphogenesis, and ion channel-mediated bioelectric patterning processes. 

We also identified non-coding regulatory signatures, including long non-coding RNAs (lncRNAs) such as MKX-AS1, NKX2-1-AS1, SIX3-AS1, ISL1-DT, LINC02849, and XIST, to name a few, which may play roles in epigenetic regulation, chromatin remodeling, or transcriptional control processes. Interestingly, protein-coding genes such as SIX3, MKX, ISL1, and NKX2-1 also emerged as top plasticity signatures across independent algorithms, validating the robustness of our causal biomarker discovery. In addition, multiple ribosomal protein genes (RPL5, RPL31, RPLP1, RPS6, RPS27, among others) were enriched in key clusters. These RPL/RPS genes, beyond their canonical roles in translation, are known to modulate cellular stress responses, p53 signaling, and post-transcriptional regulation processes, suggesting their irreducible contribution to plasticity and cell state transitions (differentiation dynamics) in glioma ecologies. 

\subsection{Neurodevelopmental Hierarchies Recapitulated in Complex Network Dynamics Suggest Lineage Bias Towards Neuronal-like Identities}

Our findings support the model that glioblastoma and DMG recapitulate embryonic developmental programmes specific to their brain region through the expression of lineage-specific markers. Prior work has shown that the putative cells of origin for IDHWT glioma align with outer radial glia (RG) stem cells in the subventricular zone (SVZ), which orchestrate early neurodevelopmental hierarchies \cite{Jessa2025,wang2025}. These stem-like cells exhibit disrupted lineage commitment and plasticity, driving tumor progression through reactivation of primitive developmental programs that impair global brain connectivity and network organization \cite{baig2024holistic,winkler2023cancer}.
Jessa et al. identified lineage-restricted transcriptional modules in pediatric gliomas using NMF and hierarchical clustering, distinguishing dorsal PAX3+/WNT-BMP OPC programs in H3.3K27M gliomas from ventral NKX6-1+/SHH OPCs in H3.1K27M subtypes \cite{jessa2019}. Our data support this dorsoventral distinction: NKX3-2 emerges as a differential marker between pediatric GBM and DMG, while NKX2-1 is shared across both subtypes (Figure~\ref{fig:bdm_nodes}). Further, NNMF-based transcriptomic modeling by Wang et al. revealed that gliomas often occupy fluid, hybrid spectrum of neurodevelopmental states that span beyond classical OPC identities \cite{wang2025}. Our findings similarly highlight bifurcation signatures pertaining to neuronal identity, neuronal progenitor and interneuron-like traits within these tumors.

Additionally, markers such as DLX5/6 and FOXR2—commonly seen in ventral telencephalon-derived interneuron lineages—suggest a shared differentiation context with FOXR2-driven neuroblastomas, which also exhibit mixed glial-neuronal phenotypes \cite{Jessa2025,wang2025}. This underscores the broader theme of disrupted developmental lineage context and epigenetic reprogramming as core drivers of glioma plasticity. In particular, DLX family genes encode transcription factors which may repress OPC fate while promoting GABAergic interneuron identities, reducing tumorigenicity in pediatric HGGs \cite{Zagozewski2015,Furst2024}. We found significant enrichment for forebrain dorsal/ventral pattern formation (GO:0021798, \( p = 3.74 \times 10^{-2} \)) among DEGs in the PID-inferred network, supporting a regional developmental architecture underlying glioma subtype specification and suggesting a promising axis for precision medicine strategies, such as cell fate reprogramming. To further support these findings, enrichment results from PID network analyses and associated transcriptional signatures are detailed in Supplementary Information (Section~\ref{si:gp_results}).

We also observed concordant signals in the bulk RNA‐seq DEGs for combined gliomas versus hNSC (Supplementary Table~\ref{tab:go_glioma_hnsc_top}), where \textit{GRIN2A}, \textit{GRIK3}, \textit{NRXN1}, \textit{SEMA6A}, \textit{ISL1}, \textit{NOTCH3}, \textit{NLGN3}, and many neurotransmitter receptors and potassium channels, were among the recurrent hits in g:Profiler-based gene enrichment analysis. The top enriched BP pathways included modulation of chemical synaptic transmission ($p_\mathrm{adj}=3.42\times10^{-19}$), regulation of trans\mbox{-}synaptic signaling ($3.42\times10^{-19}$), regulation of membrane potential ($3.96\times10^{-18}$), synapse organization ($4.99\times10^{-15}$), and axon development ($5.55\times10^{-14}$), reinforcing a neuronal/synaptic fate bias at the global transcriptome level.

\subsection{Bioelectric Patterning Genes and Neurotransmitters as Cognitive Scaffolds of Hybrid Neuronal Identity: Forecasting the Teleonomic Endpoints of Glioma Plasticity}

Furthermore, DEG analysis in Table~\ref{tab:top_deg_gliomas} reveals that glioma plasticity involves widespread downregulation of transcriptional signatures integral to bioelectric signaling and neurotransmissive control, including GRIK3, SLC24A3, KCNK10, KCNA6, SHISA6, KCNJ16, CACNG5, SLC8A3, and CHRNA9. These genes encode ion channels and transporters, including potassium (K+) channels and components of glutamate receptors, maintaining membrane potential, cell-cell and cell-matrix communication, electrophysiological patterning, and cell polarity, all of which are essential features of morphogenesis and neural lineage commitment. Their collective suppression in gliomas reflects a disruption of the bioelectric code and morphogenetic 'self' boundaries, enabling plasticity, invasion, and aberrant differentiation dynamics \cite{Levin2019}.  

Recent studies reveal that gliomas—particularly diffuse midline gliomas (DMGs) and DIPGs, as well as adult counterparts—integrate into neural circuits by forming neuron-to-glioma synapses, notably with GABAergic and glutamatergic neurons. These synaptic inputs promote malignant traits such as proliferation via depolarization-induced calcium signaling and neurotransmitter receptor signaling, representing hijacked neuroplasticity mechanisms~\cite{Barron2025, Taylor2023, Monje2020synaptic}. Our findings suggest that beyond the emergence of invasive phenotypes, this synaptic integration reflects a deeper (intrinsic) teleonomy (i.e., purposive, goal-directed behaviors) and semiotics (i.e., meaning-making processes) in glioma ecologies. They seem to suggest cell fate reprogramming toward neuron-like identities within the collective intelligence of glioma ecosystems throughout disease progression, as supported by transcriptomic enrichment of neuronal differentiation markers herein, suggesting that glioma fate cybernetics may be teleonomically biased toward neuronal functional states~\cite{winkler2023cancer, Mondal2025synaptic}. These findings position glioma plasticity regulators within a (brain) neuron-glia-immune interface, as a therapeutic window to reprogram glioma ecosystems by targeting maladaptive neurotransmission dynamics.

For instance, we identified key neurotransmitter-related genes such as GABRA5 (GABAergic signaling; downregulated DEG), GRM8 (glutamate metabotropic receptor; a BDM perturbation signature in DMG), CHRNA9 (acetylcholine), GCHFR (amine neurotransmitters and nitric oxide), and ADRA1A and ADRA2C (epinephrine signaling), indicating a broad disruption in neurotransmitter communication shaping glioma behavioural patterns. These findings suggest that bioelectric regulators may serve as cognitive scaffolds of glioma plasticity, or that at the scale of morphospace dynamics (i.e., developmental patterning), gliomas may be navigating neuronal differentiation programs albeit being 'stuck' from this lineage transition into developmentally arrested, pathological attractors. Such findings support the view of glioma ecologies as cellular identity disorders, with bifurcations into unstable, maladaptive cell fates. Some of the relevant neurotransmissive signals and neuronal circuit features identified through our causal biomarker discovery are listed in Table~\ref{tab:neuronal_signatures}. 

For instance, GABRA5, a subunit of the GABA\textsubscript{A} receptor involved in inhibitory neurotransmission, is significantly downregulated among the DEGs shown in Table~\ref{tab:top_deg_gliomas}.Furthermore, while GABRA5 polymoprhisms have been implicated with neurodevelopmental disorders and neuropsychiatric conditions, including autism spectrum and Bipolar disorder\cite{otani2005gaba,zurek2016gaba}, ADRA1A and ADRA2C encode adrenergic receptors, which have also been implicated in neuropsychiatric disorders such as schizophrenia, as susceptibility genes as epinephrine belongs to the same group of catecholamines as dopamine\cite{clark2004adra1a}. 

In contrast, GRIK3 and GRM8, are excitatory neuronal markers associated with neurodevelopmental disorders, and their emergence as differentially expressed or BDM signatures suggest a transient or hybrid neuronal identity, further highlighting the aberrant differentiation trajectories within glioma ecosystems. Furthermore, these transcriptional signatures support the emerging paradigm of systems medicine, where neuromodulators, and neurotransmitters along the neuroendocrine and brain-gut-immune axes play active cybernetic roles in steering and shaping glioma behaviors through synaptic feedback loops \cite{Krishna2023, Venkataramani2019a, Venkataramani2022, Hua2022, Taylor2023, Venkatesh2019}. Many studies have shown that brain tumors, in adults and children, such as glioblastomas exist within and exploit the complex neural circuitry of electrochemical and synaptic integration, where neuronal activity promotes tumor growth through both paracrine signaling and direct synaptic input \cite{Krishna2023, Venkataramani2019a, Venkataramani2022, Hua2022, Taylor2023, Venkatesh2019}.For instance, recent studies show that tumor-promoting GABAergic neuron-to-glioma synapses in H3K27M-altered DMGs, where GABA\textsubscript{A} receptor-mediated depolarization drives glioma proliferation—unlike in hemispheric HGGs. Clinically used drugs like lorazepam were shown to worsen DMG growth, while levetiracetam suppressed it, revealing neuroanatomical location-specific and subtype-specific neurophysiological mechanism with therapeutic implications \cite{Barron2023, Barron2025}. This opens the door for precision neuropsychiatric therapies to be repurposed for treating gliomas, offering novel, systems targeted interventions based on their neurotransmissive and neural circuit-level dysregulation. Importantly, while previous studies emphasize glioma–neuron synaptic connectomes, our presented findings reveal neurotransmitter and synaptic gene expression signatures intrinsic to gliomas themselves, embedded within their single-cell fate decisions and transcriptome-level circuitry.

Our findings also extend some findings of a recent study by Wang et al., where the authors identified a tripotent intermediate progenitor cell (Tri-IPC) in the developing human neocortex that gives rise to astrocytes, OPCs, and GABAergic interneurons \cite{wang2025neocortex}. They found that most glioblastoma (GBM) cells transcriptionally resemble Tri-IPCs, suggesting that these high-grade gliomas navigate a developmentally arrested attractor state to drive their plasticity engines. Their findings implicate Tri-IPCs, as a possible cell-of-origin and trilineage attractor for GBM, and ties gliomagenesis to disruptions in neurodevelopment, neocortical neurogenesis and gliogenesis. However, in contrast, rather than a purely GABAergic identity, our findings suggest that these gliomas exhibit a mixed, hybrid identity  with both excitatory and inhibitory neuronal signatures that may be guided toward multiple neuronal subtype bifurcations. Further, our findings suggest that the developmental morphogens expressed by these hybrid features navigate earlier embryonic patterning programs, at the mesoderm-ectoderm axis, and specifically neural tube and crest patterning processes.   

For example, DLX5/6 encode transcription factors essential for neuronal differentiation and interneuron development, while SIX3 is involved in forebrain and hypothalamic development and neuronal patterning (see Table~\ref{tab:neuronal_signatures}). This attractor state reflects a departure from purely OPC- or astrocyte-like profiles as traditionally conceived, instead revealing hybrid signatures that engage neuronal circuit specification and neuronal lineage pathways, reinforcing the hypothesis that gliomagenesis may reflect disrupted neocortical differentiation trajectories along their stalled development.

\subsection{Differentiation Therapies and Cancer Reversion as Cell Fate Reprogramming Strategies}

Our identified developmental biomarkers not only guide targeted therapies but also reveal glioma ecologies as disorders of cellular (lineage) identity, and cell fate decisions—stuck in multi-lineage, plastic states that evade terminal differentiation toward their favored lineage. This disruption of normal cell fate trajectories reflects a breakdown in teleonomic goals and systemic communication (i.e., cybernetics), where cancer hijacks and aberrantly expresses developmental programs and signaling networks for survival-based, maladaptive behaviors. Yet, we interpret that their teleonomic and semiotic trajectories hint at a favored differentiation trajectory toward neuronal-like identities, suggesting a latent developmental program. This reframes glioma ecosystems as disorders of cell fate decisions and cellular identities, where trapped in high energy, stem-like and intermediate states block transdifferentiation into neuronal lineages, thus, driving maladaptive behaviors such as aggressiveness.

The emergence of neuronal markers and activation of neuronal differentiation programs suggest that the favored lineage identity of glioma cell fates may be a navigation (trajectory) bias or commitment toward neuronal-like identities. However, their (phenotypic) epigenetic plasticity prevents stable commitment, trapping cells in higher-energy, unstable attractor states within a fluid, hybrid identity spectrum. This plasticity, while serving as an adaptability engine, results in stalled terminal differentiation along these complex attractors. Therefore, therapeutic strategies should aim to constrain this plasticity,  via targetting our identified plasticity signatures, steering cell states out of complex, transition trajectories and enabling transdifferentiation toward their favored low-energy neuronal lineage. 

We propose CRISPR/RNAi screens and pharmacological agents targeting the identified plasticity signatures, alone or in synergy, to promote glioma differentiation. These plasticity signatures may also serve as predictive features (biomarkers) that steer the glioma's cell fate transition trajectories, and hence, could be used to constrain its malignant plasticity, and emergent patterns of behaviors. Conceiving tumor ecologies as disorders of cellular identity, and cell fate decisions, suggests a psychotherapeutic-like approach to cancer care, focused on restoring systemic wholeness and redirecting maladaptive cellular behaviors toward systemic integration. As such, these plasticity markers may enable targeted reprogramming of malignant cells toward stable attractor states, promoting reintegration of fragmented tumor cell fate identities into organismal wholeness, i.e., physiologically coherent tissue ecosystems. To achieve this cell fate reprogramming, key targets based on our findings include: (1) WNT pathway activation (e.g., via GSK3$\beta$ inhibitors) combined with KDM5B inhibition, (2) GREM1 activation (CRISPRa) coupled with KDM5B inhibition (can be combined for synergetic effects with GSK3 inhibition), and (3) activation of GRIK1/3 or GABRA5 or GRM8 (CRISPRa or pharmacologically), and other identified bioelectric modulators of cell fate choices. Subtype-specific WNT activators such as WNT7A may further enhance fidelity toward neuronal reprogramming. GREM1 activation is agonistic to WNT activation, as both can synergize to promote neuroectodermal and neuronal differentiation by suppressing BMP signaling and reinforcing pro-neuronal lineage trajectories.

To promote stable lineage commitment and constrain glioma plasticity, we identify other cell fate reprogramming combinations, such as: ISL1 + NKX2-1 (CRISPRa) which may promote neuron identity; FOXD1 + GREM1 to drive neuroectodermal differentiation by suppressing TGF-$\beta$ via BMP antagonism; DLX6 + FOXG1 modulation, as they are involved in fetal forebrain development, may lock cells into terminal neuronal-like cell fates, to name a few. These neurodevelopmental regulators scored highly in our BDM analysis, and their perturbation aligns with upregulation of GABAergic markers (DLX5/6, SLC32A1, etc.) seen in our reprogrammed glioma transcriptomes. Other spporting combinations include FOXA1, FOXD1, and FOXR2 modulation, and the dual perturbation of DLX6 + GRIK3 or GRM8 or GABRA5,or some of the identified potassium channels (e.g., KCNK4, KCNK10, etc.) to name a few, for bioelectrical reprogramming. These are illustrative examples, and additional combinations of our top-ranked BDM perturbation signatures can be explored to further optimize glioma reprogramming strategies.

\subsection{Limitations}

A primary limitation of this study lies in the absence of experimental validation for our algorithmically inferred targets and biomarkers. While our neurosymbolic framework identifies key plasticity signatures, as perturbation-sensitive genes, through a combination of algorithmic complexity and network science, these predictions remain computational and require future functional validation. In particular, our pipeline provides a direct path toward experimental translation using CRISPR-based perturbation screens, enabling the empirical testing of causal regulators inferred from complexity-driven attractor dynamics. As such, prospective studies should integrate our complex dynamical systems approaches with deep learning algorithms, such as recurrent neural networks, and variational autoencoders, to enhance causal biomarker discovery and infer the complex bifurcation dynamics that steer cancer cell fate decision-making.

Another major limitation is the lack of longitudinal glioma datasets, which restricts our ability to capture temporal gene expression dynamics and evolving network states over time, to validate the predicted plasticity signatures. Access to such data would enable deeper insights into causal information flow, risk stratification, and disease trajectory modeling, thus, advancing personalized, longitudinal predictions for glioma progression and therapeutic response.

Single-cell transcriptomic analyses are also subject to inherent technical constraints, including capture inefficiencies, gene dropout, and transcriptional noise. To partially correct for these, we employed log-normalization and robust statistical approaches to account for batch effects and variability in gene expression profiles; however, some underlying artifacts may persist. Additionally, the representativeness of single-cell data is limited, potentially overlooking the influence of comorbidities, health or healthcare disparities, and social factors. Future studies should replicate our findings in larger, more inclusive and diverse cohorts to enhance translational relevance and promote accessible, and equitable, excellence-driven patient care.

The transcriptomic signatures shaping cell fate dynamics hold potential for enhancing clinical strategies, including refined disease classification, risk stratification, and forecasting treatment responses. Further, they may guide liquid biopsy-based detection of causal biomarkers for early diagnosis, tracking disease progression, recurrence, and resistance. Therefore, we advocate for the integration of multi-omic analyses, such as single-cell proteomics (e.g., CyTOF), epigenomics (e.g., chromatin accessibility, methylome profiles, Hi-C captures, etc.), metabolomics, and lipidomics, to decipher causal biomarkers across these complex, multiscale cancer processes. Assessing chromatin state and accessibility can further validate lineage or cell fate bifurcation trajectories by confirming the activity of identified transcription factors and chromatin regulators during these transitional states.

\section{Conclusions}

Algorithmic complexity–based network perturbation analyses decoded plasticity markers and disrupted communication patterns in glioma ecologies, identifying regulators of cell fate cybernetics (decisions). In combination with network science, AID measures emerged as causal discovery tools for biomarker identification, predictability, and reprogrammability of cancer cell fate. We propose that gliomas can be conceptualized as complex cybernetic systems—or maladaptive cognitive–behavioral systems—at the scale of cellular adaptive intelligence, in which fragmented cellular identities and disrupted communication networks hijack neurodevelopmental programs and malignant traits to navigate metastable or unstable phenotypic attractors, thereby driving aggressive, maladaptive behaviors \cite{Barron2023, Barron2025}. More specifically, our findings suggest that pHGGs represent disorders of cellular identity and fate decisions caused by a systemic breakdown in communication along developmental patterning processes.

Our gene regulatory network perturbation signatures highlight key neurodevelopmental and bioelectric markers—\textit{GRIK3}, \textit{ISL1}, \textit{NKX2-1}, \textit{KCNJ16}, \textit{KCNK4/10}, \textit{FOXG1}, \textit{FOXR2}, \textit{FOXA1}, \textit{FOXP1}, \textit{DLX5/6}, \textit{FOXD1}, \textit{SIX3}, \textit{SOX8}, \textit{FABP5}, \textit{WNT7A}—that influence neuronal differentiation programs and are predicted to control glioma fate decisions and malignant plasticity. Neurotransmitter-related genes such as \textit{GABRA5}, \textit{CHRNA9}, \textit{GRM8}, \textit{ADRA1A}, and \textit{ADRA2C} implicate dysregulated neural circuit integration, neurochemical signaling, and hybrid neuronal differentiation with mixed lineage cues in steering glioma cell fate. Such systems-level targeted therapies could also modulate physiological cybernetics—including brain–gut–immune and neuroendocrine axes—and warrant validation through CRISPR screens and functional assays.

DIPG (DMG) displayed strong BDM signatures for neuronal differentiation regulators (\textit{DLX5/6}, \textit{SIX3}, \textit{FOXG1}) and OPC-like markers (\textit{OLIG1/2}), suggesting a hybrid identity biased toward forebrain developmental programs and neuronal-like fate commitment. These genes indicate that gliomas hijack fetal neurodevelopmental programs through disrupted communication networks in patterning processes. In contrast, IDHWT GBM exhibited BDM signatures associated with neural stem cell–like plasticity and neuroimmune/metabolic rewiring, with key roles for \textit{FOXD1}, \textit{GREM1}, \textit{CD24}, and TNF/cytokine-driven niche construction.

Developmental transcription factors including the FOX family, \textit{NKX2-1}, \textit{DLX5/6}, and \textit{WNT7A} emerge as strong candidates for precision oncology strategies based on cell fate reprogramming and tumor ecosystem engineering—encompassing cancer reversion, precision gene editing, and differentiation therapies, as they orchestrate lineage specification and fate transitions via morphogenetic patterning processes. Their dysregulation reveals how tumors exploit both bioelectric signaling and developmental plasticity genes to sustain malignant behaviors. Our findings also point to a teleonomic disruption—a systemic breakdown in morphogenesis and ecological cybernetics between cell fates and their environments—leading to developmental arrest (i.e., stalled differentiation) and destabilized fate identities, with plasticity serving as the cybernetic scaffold for re-steering toward stable, integrated outcomes.

The discovered cell fate transition markers implicates bioelectric/neurotransmitter circuitry (HCN1, KCNA6, GRIK and GRIN-like; GABRA5, ADRA1A, etc.) and ECM–chemokine coupling (CHST15–ZIM2; CXCR4–NAP1L2) as actionable levers, suggesting membrane-potential modulation or CXCR4–ECM pathway interference could disrupt neuron–glioma synapses and immune–neuro interfaces.

To our advantage, these perturbation signatures and network biomarkers may represent therapeutic vulnerabilities, analogous to how \textit{FOXA2} (with HDAC2) reprogrammed colorectal cancer cells into stable, non-malignant phenotypes \cite{gong2025}. Leveraging graph network–based complexity signatures—interpretable both as drivers of glioma invasiveness and adaptive evolution, and as predictive features for forecasting progression—we propose that plasticity signatures can serve as clinically relevant prognosticators and early recurrence diagnostics via liquid biopsy and multi-omics profiling.

To conclude, we propose that the identified BDM network signatures (plasticity markers) could be deployed individually or in combination as differentiation therapies \cite{aguade2022transition}, reprogramming glioma cells toward non-proliferative, non-invasive fates. Translational validation through CRISPR screens, functional assays, and perturbation studies in preclinical models is essential to advance these strategies. This systems medicine approach frames healing as the reintegration of maladaptive cellular states into a coherent physiological whole—where precision oncology strategies for cell fate reprogramming constrain malignant plasticity and steer tumor collective behaviors toward stable identities, fostering wholeness and enabling safer, targeted therapies in patient-centered care.

\begin{table}[H]
\centering
\caption{Synaptic/Neuronal Lineage and Circuit Specification Gene Signatures Discovered in our Analyses.}
\label{tab:neuronal_signatures}
\begin{tabular}{l|p{11cm}}
\textbf{Gene} & \textbf{Function in Synaptic and Neuronal Lineage Specification} \\
\hline
\textbf{GRIK3} & Glutamate receptor; involved in synaptic transmission. \\
\hline
\textbf{DLX5/6} & Key transcription factors for GABAergic interneuron development. \\
\hline
\textbf{SIX3} & Involved in forebrain and hypothalamic development, neuronal patterning. \\
\hline
\textbf{ISL1} & LIM-homeodomain gene involved in motor neuron differentiation. \\
\hline
\textbf{GABRA5} & GABA-A receptor subunit; marker of GABAergic signaling. \\
\hline
\textbf{CHRNA9} & Nicotinic acetylcholine receptor; modulates neuronal excitability. \\
\hline
\textbf{GRM8} & Metabotropic glutamate receptor; modulates excitatory synaptic transmission and neuronal signaling. \\
\hline
\textbf{ADRA1A} & Alpha-1 adrenergic receptor; neurotransmitter-related function in CNS. \\
\hline
\textbf{FOXG1, FOXP1, FOXD1} & Transcription factors involved in neuronal development and fate specification. \\
\hline
\textbf{KCNK4/10} & Potassium channels involved in setting neuronal membrane potential. \\
\hline
\textbf{SHISA6} & Involved in excitatory, glutamatergic synapses and ionotropic glutamate receptor signaling. \\
\end{tabular}
\end{table}

\section*{Data and Code Availability}

All code and processed results supporting this study are publicly available, and can be accessed at:
\url{https://github.com/Abicumaran/Glioma_Network_BDM_Signatures}.

\bibliographystyle{unsrt}  
\bibliography{references}

\clearpage
\appendix
\section{Supplementary Information}

\subsection*{Glioma Ecology: Context and Epigenetic Plasticity}

Glioblastoma and DMGs are spatially distinct pHGG subtypes, with debated differences in their cells of origin\cite{wang2025}. DMGs are predominantly caused by H3.1/2 (canonical) pontine/brainstem (harboring ACVR1 mutations) and H3.3 (noncanonical) thalamic K27M histone mutations, which impair K27 trimethylation, a repressive chromatin mark, by the Polycomb repressor complex 2 (PRC2). This oncohistone mutation affects genes encoding two isoforms of the Histone H3 protein, H3F3A (H3.3) and HIST1H3B (H3.1), and acts as a key epigenetic driver in DMGs  \cite{schwartzentruber2012,ocasio2023}. It is identified in up to 80\% of pediatric cases, contributing to tumor progression through dysregulated chromatin modification, and gene expression dynamics \cite{Saratsis2024}. The global loss of H3K27me3 marks across developmental genes results in the aberrant expression of morphogenetic programs, and hence, phenotypic plasticity as an emergent, adaptive trait \cite{ocasio2023, schwartzentruber2012, khuongquang2012}. 

Importantly, following the 2021 WHO CNS classification, the term glioblastoma is now primarily reserved for IDH-wildtype grade 4 astrocytoma \cite{louis2021who, alnahhas2024molecular}. This designation is distinct from DMGs which primarily affect children \cite{louis2021who, alnahhas2024molecular}. Although the term DMG has replaced earlier classifications such as DIPG that may have grouped such tumors under glioblastoma, in this study we continue to use the term glioblastoma for samples derived from Neftel et al. (2019)\cite{neftel2019integrative}
, and the term DIPG for K27M mutant DMG samples obtained from the dataset from Filbin et al. (2018)\cite{filbin2018developmental}, consistent with the terminology used at the time. 

Traditionally, H3K27M HGGs, such as most pediatric DMGS, were presumed to originate from oligodendrocyte precursor cells (OPCs) as their cell-of-origin (CO), with context-dependent signaling pathways—dorsal (thalamic) for H3.3 and ventral for H3.1 \cite{jessa2019, jessa2022, liu2022}. Meanwhile, glioblastomas, more common in adults, can be IDH1/2-mutant (typically progressing from lower-grade gliomas) or predominantly IDH-wildtype. Recent findings using machine learning algorithms demonstrate that neural stem cells (NSCs) from the subventricular zone (SVZ), such as radial glia, are the root CO for IDHWT gliomas, although differences in their neurodevelopmental patterning processes and developmental networks from DMGS, remain an open question \cite{mathur2024, wang2025}. Our previous findings also supported this view, demonstrating that glioma stemness does not conform to a fixed hierarchical model but rather reflects a dynamic, feedback-regulated continuum of stem-like and progenitor-like plastic states embedded within nonlinear attractor landscapes \cite{Uthamacumaran2025}.

Although glioblastoma and H3K27M-mutant DMG differ in molecular features, and anatomical localization—glioblastoma typically arising in the cerebral cortex and DMG in the brainstem or midline structures—emerging evidence suggests overlapping expression of neural stem cell (NSC) lineage markers, and hybrid cellular identities favoring a neuronal-like state \cite{wang2025, Uthamacumaran2025}, suggesting causal network processes steering their plasticity. This indicates that both PHGG subtypes may be navigating similar neurodevelopmental differentiation programs, a blockade of which may cause a developmental arrest in a plastic, hybrid identity with a bias towards neuronal lineage transition\cite{roberts2023, Larsson2024, wang2025}. In other words, we suggest that these malignant cell fates exhibit maladaptive emergent behaviors—such as aggressiveness and therapy evasion—by remaining trapped in early progenitor programs, unable to complete differentiation or transdifferentiation towards their favored neuronal lineage commitment, due to unconstrained plasticity (i.e., high energy states on the attractor landscape) \cite{uthamacumaran2022alg,uthamacumaranzenil2022,Uthamacumaran2025}.

In return, this leads to lineage-specific developmental stalling, maintained by chromatin deregulation and aberrant signaling dynamics \cite{Chen2024DevOrigins}. This suggests their teleonomy—or goal-directed differentiation—is arrested within pathological stem-like or progenitor states, leading to disrupted meaning-making processes (Semiotics; such as aberrant signaling), and 'maladaptive' behavioral patterns such as invasion, heterogeneity, and aggressiveness, as a result of the systemic breakdown of cellular communication and cell fate cybernetics.  

Thus, the purpose of this study is to investigate how DIPG (DMG) and pGBM gliomas differ from healthy developmental brains at the level of RNA transcriptomic state-space by use of algorithmic network complexity measures \cite{zenil2019}. To the best of our knowledge, this is the first application of algorithmic complexity-based perturbation analysis to decode the causal trajectories underlying cell fate choices across pHGG subtypes \cite{zenil2019, wang2025}. This approach enables the identification of plasticity markers that govern cell fate dynamics and the reconstruction of differentiation trajectories, providing a framework to reverse-engineer and reprogram glioma cell fates toward transdifferentiation or terminal differentiation into non-malignant states~\cite{gong2025, xiang2025}.

This complex systems approach quantifies the emergent behaviours of collective cell fate decisions within glioma networks by inferring the potential attractors they navigate in gene expression (network) state-space, revealing dynamical systems insights into their developmental trajectories and pathological differentiation states~\cite{levin2021,barcenas2024}. Unlike traditional statistical approaches, algorithmic complexity-based perturbation analysis—namely, algorithmic information dynamics (AID)—explores how local interventions affect global topological stability and the emergent behavioural repertoire of the network~\cite{Zenil_Causal2019, Zenil_Decomp2018}. Under the AID framework, algorithmic information theory links algorithmic complexity to the generative rules of a system, quantifying how simple or random-like the shortest program is that constructs its structural irregularities and patterns of behavior. This enables decoding the underlying causal mechanisms and emergent principles driving complex system dynamics. When combined with perturbation analysis, it allows the prediction of complex dynamics, such as local state-transitions, and collective cell fate trajectories \cite{uthamacumaranzenil2022}.

By use of algorithmic complexity measures such as the Block Decomposition Method (BDM), and lossless compression algorithms like Lempel-Ziv-Welch (LZW), rather than purely probabilistic entropy or correlational analyses, and traditional stochastic models or reductionist statistical approaches within the framework of 'pseudotime ordering' methods, AID captures the causal information dynamics and structural regularities underlying cell state transitions ~\cite{Zenil_Causal2019, Zenil_Decomp2018}. Further, it decodes the plasticity signatures steering these complex attractor dynamics as computational processes. Perturbation analyses across this attractor landscape reveal which network substructures, such as nodes, links, or modules exert maximal control over network dynamics, thus identifying potential biomarkers or intervention 'tipping points' to steer collective cell fate behaviors, such as plasticity, therapy resistance, and stalled cell fate commitments ~\cite{Zenil_Causal2019, Zenil_Decomp2018}. These perturbed attractor dynamics represent navigable critical points in a high-dimensional decision space, in silico, offering network perturbation signatures as potential targeted therapies and precision biomarkers steering collective cell fate choices in cancer ecosystems. Further, it provides a computational medicine framework to decode graph-theoretic signatures and plasticity markers as drivers of collective cellular agency, maladaptive behaviors, and reprogrammable identity dynamics in glioma ecologies. 

Rather than fixed identities, this study proposes that gliomas explore a neurodevelopmental continuum—of hybrid cellular identities—with sensitive dependence on initial conditions (cell of origin).  Specifically, our findings reveal alterations in morphogenetic processes, neurotransmission pathways, and electrochemical signaling related to neuronal differentiation, highlighting disrupted neurodevelopmental programs in pediatric gliomas \cite{liu2024}. These findings challenge assumptions about cellular identity and origin, raising critical questions about neurodevelopmental hierarchy, lineage specification, and the complex networks that steer the pathological attractors and disrupted teleonomic (goal-directed) behavioral patterns of tumor ecosystems \cite{baig2024holistic, winkler2023cancer}. Pathological attractors reflect disrupted teleonomy in cell fate cybernetics (i.e., communication and decision-making), as a result of which maladaptive behaviors emerge, driven by aberrant semiotics—misinterpreted or dysregulated meaning-making processes within the signaling and regulatory networks of cellular identity. These complexity-driven network biomarkers may serve as putative therapeutic targets for cancer cell fate modulation  by altering lineage-specific differentiation programs and hence, cell fate plasticity networks\cite{sole2021, xiang2025}. 

\subsection*{Plasticity Networks and Pathological Attractor Dynamics}

Phenotypic plasticity, often termed developmental or epigenetic plasticity depending on the scale of interaction (e.g., morphospace-level patterning), or as fate plasticity and lineage plasticity at subpopulation dynamics (i.e., cell fate vs. lineage identity decisions), will here be used as an umbrella term to denote the collective behavioral patterns of these complex dynamical systems denoting phenotypic switching or cell fate transitions\cite{Hanahan2022}. 

The stem-like identity in gliomas remains debated, with some proposing glioma stem cells as a distinct subpopulation, while others reject a fixed, binary stem cell hierarchy in favor of a continuum of plastic identities, including progenitor-like identities bifurcating from neural stem cells (NSC), i.e., intermediate states at the edge of criticality between stemness and differentiation, with multilineage bifurcations \cite{desisto2025prmt5, hemmati2003cancerous, daveiga2022glioma, poorva2025nk}. Our previous studies found that pHGG cells navigate a pathological attractor  enriched with NSC signatures, characterized by multilineage potential, high entropy, and algorithmic complexity, reflecting an unstable state with complex bifurcation dynamics\cite{uthamacumaran2022alg,uthamacumaranzenil2022,Uthamacumaran2025}. By attractors, we refer to the emergent structures representing the long-term behavioral patterns that steer cell fate decisions and differentiation trajectories within the state space of transcriptional regulatory networks \cite{uthamacumaran2022alg,uthamacumaranzenil2022,Uthamacumaran2025}. These attractors are shaped by 'plasticity networks', where bifurcation signatures and critical tipping points mark state-transitions in transcriptomic state-space, steering cell fate trajectories toward distinct lineage identities or causing them to stall in high-entropy, plastic states\cite{uthamacumaran2022alg,uthamacumaranzenil2022,Uthamacumaran2025}. Predicting these patterns of cell fate transitions enables the forecasting of lineage biases or fate preferences that glioma cells tend to follow post-bifurcation, offering actionable targets for cellular reprogramming strategies to overcome developmental arrest and redirect cells toward  stable cell fate identities\cite{uthamacumaran2022alg,uthamacumaranzenil2022,Uthamacumaran2025}. 

In evidence to the fate reprogrammability of these plastic ecosystems, in our previous findings, we found a fluid identity between stem-like and neural progenitor-like states, with a decentralized distribution of phenotypes across the tumor's developmental (Waddington) landscape. The Waddington landscape is an attractor-based view of gene-regulatory dynamics, wherein valleys (basins) are stable cell-fate phenotypes, ridges are barriers, and cell fate bifurcations denote lineage transitions or aberrant states, such as tumor development, across morphospace \cite{Waddington1957,Huang2012}. We showed that gliomas exhibit neural stem cell (NSC)-like identities and bifurcate into neuronal progenitor (NPC), oligodendrocyte progenitor (OPC), and mesenchymal (MES)-like lineages, with fluid, hybrid identities reflecting collective tumor plasticity\cite{Uthamacumaran2025}. NPC-like fates are intermediate progenitor states derived from NSCs, capable of differentiating into neurons or astroglial fates, whereas OPC-like states are more lineage-restricted\cite{Uthamacumaran2025}. Within the MES-like bifurcating lineage, we also identified a myeloid–hematopoietic stem cell (HSC)-like population, indicative of immune mimicry, immune evasion, and developmental plasticity, i.e., malignant traits and predictive features linked to the emergence of invasive, aggressive, and therapy-resistant glioma phenotypes.

The study further revealed a (lineage) cell fate bias toward NPC-like identities, enriched for neuronal differentiation markers, as a favored trajectory. This suggests that malignant cell fates may remain stalled in plastic, high-energy states or gravitate toward the multilineage NPC state due to developmental constraints within the teleonomic structure of the attractor landscape\cite{uthamacumaran2022alg,uthamacumaranzenil2022,Uthamacumaran2025}. This neuronal fate preference was supported by pseudotime trajectory inference, enrichment of neuronal differentiation markers such as ASCL1, SOX11, and DCX, and entropy-based attractor dynamics, which collectively indicated greater prevalence of NPC-like states compared with other lineages\cite{uthamacumaran2022alg,uthamacumaranzenil2022,Uthamacumaran2025}.Thus,we predicted that NPC-lineage bias may reflect glioma cell fate plasticity along a fluid, hybrid spectrum, while also revealing a therapeutic vulnerability for cell fate reprogramming strategies such as promoting neuronal-like transdifferentiation.

Phenotypic plasticity acts as the 'creative' scaffold or "evolvability engine" of glioma ecologies, steering these plastic cell fates towards malignant traits and maladaptive behaviors, such as (intratumoral) cellular heterogeneity, therapy resistance, tumor recurrence, and immune evasion, as cells navigate the adaptive continuum from plastic to invasive identities in response to their environmental perturbations\cite{zenil2019,Hanahan2022,uthamacumaran2022alg, uthamacumaranzenil2022, Uthamacumaran2025}. In essence, plasticity reflects the adaptive creativity of complex systems, enabling novelty generation ( of phenotypic states) to navigate and withstand evolutionary selection pressures. In the context of attractor dynamics, plasticity signatures, or transition markers, correspond to bifurcation signatures along cell fate differentiation trajectories which regulate or drive phenotypic plasticity. In our previous studies, we observed a trilineage attractor with a pitchfork bifurcation, marking critical transitions across NPC-like, OPC-like, and HSC/MES-like lineages, with a lineage bias towards NPC-like states enriched with neuronal differentiation signatures\cite{Uthamacumaran2025}. This trilineage attractor may recapitulate the tri-potent NSCs of a healthy developing brain, which can differentiate into neurons, astrocytes, and OPCs (oligodendrocytes) \cite{liu2012developmental,laurenge2023cell,huang2021cracking}.

\subsection*{Cancer as a Systemic Disorder of Communication Among Cellular Collectives: A Complex Systems Approach to Cell Fate Behaviors}

We outline a theoretical framework grounded in complex systems theory—the study of emergent, collective behaviors—through which tumors are viewed as dynamic ecologies exhibiting collective intelligence, irreducible from their interactions with the microenvironment. Gliomas can be fundamentally understood as a disruption of cellular communication networks, not only between individual cells, but across multiscale, hierarchical layers of biological organization, from genes to tissues to the surrounding social networks in the tumor microenvironment \cite{levin2021, mcmillen2024collective}. Morphogenesis (pattern formation) is a tightly coordinated developmental process governed by a collective intelligence of cells responding to spatial constraints and temporal gradients of signaling, gene regulation, and multiscale feedback loops with teleonomic (goal-directed) behaviors, i.e., cell fate specification and differentiation \cite{mcmillen2024collective}. The disruption of this orchestrated developmental control, and differentiation hierarchy—whether at the level of transcriptional circuits, bioelectric gradients, or extracellular cues—can result in abnormal or stalled cell fate decisions, and disrupted (or in the case of gliomas, hybrid high-energy intermediates of lineage identities) laying the 'initial conditions' for driving malignant behaviors in tumor ecologies. Through this lens, cancer cell fate control becomes a problem of restoring lost integration with the organism's physiological and developmental systems. Disrupted epigenetic processes and gene expression programs underlying morphogenesis (pattern formation) prevent cancer cells from terminally differentiating and reintegrating into normal tissue hierarchies, effectively rendering cancer a systemic and developmental communication disorder, and a decentralized intelligence fragmented from these modular hierarchies \cite{mcmillen2024collective}. 

Cancer cell fate control involves steering collective cellular behaviors by modulating feedback loops and regulatory network circuits \cite{gong2025}. Network analysis can reveal key control nodes whose perturbation enables reprogramming of cell fate trajectories. For example, the Goodwin oscillator illustrates how simple feedback generates temporal gene expression oscillations, modeling the nonlinear epigenetic dynamics underlying cell plasticity \cite{goodwin1963temporal, goodwin1965oscillatory}. Integrating such dynamics with cybernetic control allows timed therapeutic interventions aligned with intrinsic cellular rhythms \cite{goldberg2007epigenetics}, mapping onto Waddington’s epigenetic landscape where bifurcations and attractors shape developmental trajectories \cite{noble2015waddington, ferrell2012bistability}.

Emerging multiomics studies in glioma biology increasingly reveal lineage bifurcations and patterned transitions in gene expression, suggesting that underlying causal attractors may be steering cells toward specific phenotypic states—supporting dynamic attractor models as a powerful explanatory framework \cite{Larsson2024, Jessa2025, wang2025}. Traditional clustering methods, including manifold learning and dimensionality reduction, often fail to capture such latent causal structures or temporal dynamics. In contrast, attractor-based models rooted in dynamical systems theory explain emergent patterns of collective behaviors—such as complex nonlinear oscillations—that shape the stability or plasticity of cellular identities during cell state transitions\cite{zenil2019, levin2021}. These attractors are inferred from network topology, where gene or protein interactions define cell fate behaviors \cite{sole2021, zenil2019}. This necessitates a complex systems framework capable of tracing how individual cell fate decisions give rise to (emergent) collective dynamics, while inferring underlying attractor state transitions from a given matrix or graph network topology.

As such, AID integrates tools from computer science and complexity theory to uncover causal structures within Gene Regulatory Networks (GRNs). Central to AID is Kolmogorov complexity, defined as the shortest algorithmic description of a dataset, such as a gene interaction network or adjacency matrix \cite{zenil2019}. Unlike statistical entropy measures that rely on assumed distributions, algorithmic complexity captures deep structure and pattern regularity—even in noisy or aperiodic systems. Moreover, in resonance with Schmidhuber’s theory of intelligence as compression, we interpret tumor ecology's collective behaviors as a form of cellular cognition—where cancer cells adapt to microenvironmental constraints by optimizing internal models and patterns of behaviors that maximize their survival \cite{schmidhuber2009compression}. In this view, tumor ecologies act as cognitive-perceptual systems, or collective intelligences, seeking algorithmic compressibility in their signaling dynamics, and AID tools such as BDM and compression scores can directly verify these predictions. 

AID provides algorithmic, graph-theoretic measures of network complexity that capture the underlying complex dynamics of gene regulatory systems, offering a robust dynamical systems measure of causal information. This enables the prediction and forecasting of  cell state transition dynamics, and plasticity signatures from single-cell gene expression patterns by revealing attractor-driven, collective cell fate decisions \cite{zenil2019, uthamacumaran2022alg, uthamacumaranzenil2022, Uthamacumaran2025}.

Thereby, our computational medicine framework reframes cancer as a complex cybernetic system marked by a breakdown in systemic communication during pattern formation (morphogenesis) and developmental processes. The inability of cancer cells to reach terminal differentiation and reintegrate into physiological morphospaces (i.e., pattern formation space) reflects a dysregulation of intercellular dialogue during pattern formation, as revealed by the identified developmental and morphogenetic signatures in our causal biomarker discovery. Disrupted cell-to-cell and cell-to-matrix signaling severs feedback between tumors and the broader organismal system, giving rise to maladaptive attractor states. 

By leveraging algorithmic complexity and network science, we can identify predictive biomarkers for early recurrence detection or therapy resistance, and design targeted interventions that reprogram cellular collectives toward stable attractor states. This approach advances patient-centered precision medicine—not only grounded in multiomics, but also in restoring physiological coherence through cybernetic integration of cellular signaling networks\cite{Larsson2024, zenil2019}. Insights into tumor developmental dynamics and cell fate decision-making can guide precision therapies, as AID provides a computational systems framework for decoding glioma network state-space, and understanding phenotypic plasticity—the evolvability engine driving emergent behavioral patterns such as aggressiveness and therapy resistance in tumor ecosystems \cite{mehta2024, shin2023}. Such an approach allows us to decode the causal information dynamics steering pathological attractors in collective cell fate choices within cancer networks. 

\subsubsection*{Glioma Ecologies as Identity Disorders of Cell Fate Decisions}

Based on our findings, we propose that gliomas may be conceptualized as maladaptive cognitive-behavioural systems operating at the level of cellular collective intelligence, driven by fragmented cellular identities that split or bifurcate along a hybrid, fluid spectrum of phenotypic states due to disrupted developmental (morphogenetic) programs and aberrant synaptic-like signaling circuits. The disrupted plasticity networks identified herein may represent viable targets for cellular reprogramming strategies, such as (trans)differentiation or cancer reversion therapies, drawing parallels to neuroplasticity-based psychiatric therapies aimed at restoring behavioral homeostasis.

This view supports the emerging perspective of cancer ecologies as cognitive-perceptual systems and dysregulation of collective cellular intelligence, wherein tumors represent disorders of cellular identity and cell fate decisions. As such, gliomas function as complex cybernetic systems characterized by aberrant signaling dynamics and systemic breakdowns in cellular communication. By predicting, and modulating these aberrant ecological signals—including cell adhesion molecules, bioelectric cues, transcription factors, and developmental morphogens that act as plasticity signatures—we can strategically hijack these plasticity networks to reprogram malignant cell fates and restore fragmented cellular identities toward physiological coherence and systemic integration.
\newpage

\subsection*{Algorithmic Complexity- Based Perturbation Signatures}

\begin{figure}[h!]
\centering
\includegraphics[width=0.7\textwidth]{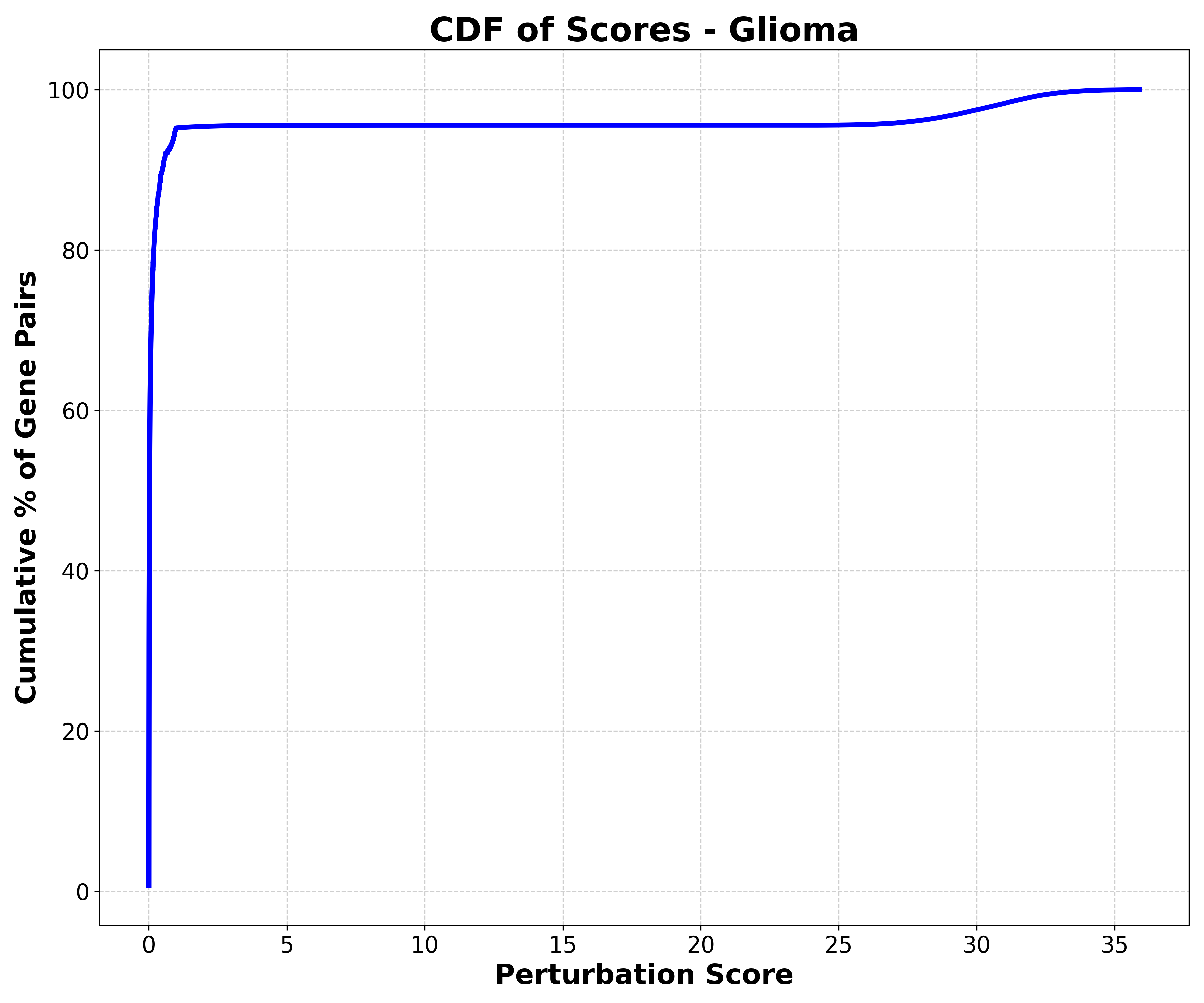}
\caption{CDF of the DEG Spearman Network for Glioma Combined. This plot shows the cumulative distribution of absolute Spearman correlation BDM perturbation scores for gene-gene links in the Combined Glioma network.}
\label{fig:cdf_glioma}
\end{figure}

\textbf{Explanation of Cumulative Distribution Function (CDF) Plots for Link (Gene pair)Perturbations from Spearman metric-derived DEG Adjacency Matrices:}  
The \textbf{X-axis} represents perturbation scores, ranging from low to high.  
The \textbf{Y-axis} indicates the cumulative percentage of DEG (genes) with BDM shift perturbation scores less than or equal to a given value. The cumulative distribution frequency (CDF) plots highlight the distribution of link (edge) perturbation scores and show what proportion of genes fall below specific thresholds.  

\begin{table}[h!]
\centering
\caption{DIPG Gene Pair (Link) Spearman DEG Network BDM Perturbations}
\begin{tabular}{@{}lll@{}}
\toprule
\textbf{Gene 1} & \textbf{Gene 2} & \textbf{BDM (abs)} \\ \midrule
C1orf116 & ADAMTS6 & 35.8301 \\
FRMPD2B & MIG7    & 35.8243 \\
MIG7    & FRMPD2B & 35.8243 \\
SLC30A4 & MDFI    & 35.8135 \\
MDFI    & SLC30A4 & 35.8135 \\
ARSD    & DLGAP2  & 35.7862 \\
DLGAP2  & ARSD    & 35.7862 \\
ARSD    & PRIMA1  & 35.7598 \\
ARSD    & SAMSN1  & 35.7598 \\
PRIMA1  & ARSD    & 35.7598 \\
SAMSN1  & ARSD    & 35.7598 \\
FUT9    & NXPH3   & 35.7598 \\
ZNF626  & TXNDC2  & 35.7598 \\
TXNDC2  & ZNF626  & 35.7598 \\
GCAWKR  & FLRT1   & 35.7598 \\
FLRT1   & GCAWKR  & 35.7598 \\ \bottomrule
\end{tabular}
\end{table}

\begin{table}[h!]
\centering
\caption{GBM Gene Pair (Link) Perturbations. WT1 (Wilm's Tumor protein) intersected as a transcription with most of these genes.}
\begin{tabular}{@{}lll@{}}
\toprule
\textbf{Gene 1} & \textbf{Gene 2} & \textbf{BDM Value} \\ \midrule
TDRD7    & ACHE       & 35.9557 \\
ZNF626   & AQP7P3     & 35.9557 \\
PPP2R5B  & TMEM35A    & 35.9557 \\
RRP9     & PINLYP     & 35.9557 \\
CLDN11   & ABCA9      & 35.9557 \\
MEIS3    & NDUFA4L2   & 35.9557 \\
LOC105378954 & HLA-DRA & 35.9557 \\
NDUFA4L2 & MEIS3      & 35.9557 \\
ABCA9    & CLDN11     & 35.9557 \\
NID2     & AZIN2      & 35.9557 \\
AMH      & KCNMB1     & 35.9557 \\
AZIN2    & NID2       & 35.9557 \\
FGF7     & EDIL3-DT   & 35.9557 \\
TMEM191B & ARMCX3     & 35.9557 \\
PODN     & RAB33A     & 35.9557 \\
LRRN1    & ADRA1D     & 35.9557 \\
ARMCX3   & TMEM191B   & 35.9557 \\
TMEM35A  & PPP2R5B    & 35.9557 \\
TRIM69   & TRABD2A    & 35.9557 \\
RAB33A   & PODN       & 35.9557 \\
KCNMB1   & AMH        & 35.9557 \\
AQP7P3   & ZNF626     & 35.9557 \\ \bottomrule
\end{tabular}
\end{table}

\begin{table}[h!]
\centering
\caption{Gliomas Combined Gene Pair (Link) Perturbations}
\begin{tabular}{@{}lll@{}}
\toprule
\textbf{Gene 1} & \textbf{Gene 2} & \textbf{BDM Value} \\ \midrule
ACSM4    & KIAA2012  & 35.8973 \\
BCAM     & TMEM150C  & 35.8529 \\
BEND5    & PEG13     & 35.8529 \\
SQOR     & GGT2P     & 35.8529 \\
SQOR     & EVC2      & 35.8529 \\
CACNA2D4 & RELN      & 35.8529 \\
LARP6    & EVC2      & 35.8529 \\
SBSN     & CCL28     & 35.8529 \\
TMEM150C & BCAM      & 35.8529 \\
P2RY6    & HLA-F     & 35.8529 \\ \bottomrule
\end{tabular}
\end{table}

Pyrimidine metabolism is linked to therapy resistance. Most of the perturbation links are associated with metabolic rewiring mitochondrial energy processes in the combined glioma network analysis. 

\begin{table}[h!]
\centering
\caption{Top Centrality Genes for GBM PID Network in DEG space. Most of these gene signatures were associated with the transcription factor EGR1 in g:Profiler analysis.
Genes like EGR1, a transcription factor, link this network to epigenetic regulation and tumor progression. Neurodevelopmental roles are suggested by HCN1 and ADRA1A, which influence neuronal signaling. Metabolic processes are highlighted by TRAP1 and SLC2A6, regulating mitochondrial and glucose transport. Immune modulation is evident from TNFRSF8 and GBP1P1, supporting a tumor microenvironment with neuro-glial lineage plasticity favoring immune evasion and metabolic adaptability.}
\begin{tabular}{ll}
\toprule
\textbf{Betweenness/Closeness} & \textbf{Eigenvector/Hub Score} \\
\midrule
SNHG4 & FRK \\
CD24 & GAD2 \\
TRAP1 & SLC51A \\
SVEP1 & MATN2 \\
PTPN3 & SVEP1 \\
ZNF440 & PIGZ \\
HCN1 & TGFB3 \\
TDRD7 & TMEM45A \\
FEZ1 & CPZ \\
SAMD9L & SLC2A6 \\
DDN & ADAMTS12 \\
ADRA1A & GBP1P1 \\
LAMP5 & \\
SLC46A3 & \\
TNFRSF8 & \\
TOP1MT & \\
FYN & \\
IFIT2 & \\
WHAMMP3 & \\
ZNF347 & \\
\bottomrule
\end{tabular}
\end{table}

\begin{table}[h!]
\centering
\caption{Top Centrality Genes for DIPG on PID Network DEGs. Centrality genes like DLX5, DLX6, ZIC1, and SIX3 are neurodevelopmental regulators critical for neuronal differentiation, highlighting a neuronal lineage preference in DIPG. HAPLN2 and NAV2 suggest roles in extracellular matrix remodeling, cell adhesion, neuronal migration, and axonal guidance, supporting tumor progression and neuro-glial interactions. LITAF links to immune modulation, while PRUNE2 and ZNF320 may have epigenetic regulatory roles. These genes indicate a neuro-glial progenitor cell of origin with a bias toward a cellular identity favoring neuronal differentiation and plasticity.}

\begin{tabular}{ll}
\toprule
\textbf{Betweenness/Closeness} & \textbf{Eigenvector/Hub Score} \\
\midrule
H4C6 & HAPLN2 \\
DLX6 & STXBP5L \\
NKX2-1 & FHIP1A-DT \\
PITX2 & NACAD \\
NAV2 & ACHE \\
FYN & SOX21-AS1 \\
ZIC4 & PRUNE2 \\
UBE2E2 & SNX10 \\
ZIC1 & REEP1 \\
ADD3 & VWF \\
LITAF & GNG8 \\
DLX5 & B3GALNT1 \\
DCHS1 & ZNF833P \\
FHL1 & HHIP \\
ZNF320 & HPCA \\
SIX3 & MTSS1 \\
\bottomrule
\end{tabular}
\end{table}

\begin{table}[h]
\centering
\caption{Top Spearman BDM for single-cell IDHWT.}
\begin{tabular}{ll}
\hline
\textbf{Node Perturbation Genes} & \textbf{Link Perturbation Genes} \\
\hline
CD74                             & MTRNR2L8-MTRNR2L1               \\
RPL4                             & MTRNR2L10-MTRNR2L6              \\
RPS6                             & MTRNR2L2-MTRNR2L1               \\
NPC2                             & MTRNR2L8-MTRNR2L10              \\
HLA-DRA                          & MTRNR2L1-MTRNR2L10              \\
S100A11                          & MTRNR2L2-MTRNR2L8               \\
HLA-E                            & MTRNR2L8-MTRNR2L6               \\
SRGN                             & MTRNR2L1-MTRNR2L6               \\
LAPTM5                           & TYROBP-FCER1G                   \\
FCER1G                           & HLA-DPA1-TYROBP                 \\
RPS3A                            & HLA-DPA1-HLA-DRB1               \\
HLA-DRB1                         & HLA-DRB1-TYROBP                 \\
HLA-DPA1                         & HLA-DRA-HLA-DRB1                \\
RPS4X                            & LAPTM5-TYROBP                   \\
B2M                              & MTRNR2L2-MTRNR2L10              \\
HLA-C                            & HLA-DPA1-FCER1G                 \\
RPS27                            & TYROBP-HLA-DMB                  \\
RPL9                             & HLA-DRB1-FCER1G                 \\
RPL17                            & MTRNR2L2-MTRNR2L6               \\
PCDHGC3                          & HLA-DRA-TYROBP                  \\
HLA-B                            & HLA-DPA1-LAPTM5                 \\
LHFPL3                           & HLA-DPA1-HLA-DMB                \\
VIM                              & LAPTM5-FCER1G                   \\
SPARC                            & HLA-DRB1-HLA-DMB                \\
EEF1G                            & HLA-DRA-HLA-DPA1                \\
ARHGDIB                          & TYROBP-ARHGDIB                  \\
JUN                              & LAPTM5-HLA-DRB1                 \\
EGR1                             & HLA-DRA-FCER1G                  \\
PTN                              & FCER1G-HLA-DMB                  \\
RPL5                             & HLA-DRA-HLA-DMB                 \\
PDGFRA                           & LAPTM5-ARHGDIB                  \\
S100A16                          & FCER1G-ARHGDIB                  \\
CD63                             & LAPTM5-HLA-DMB                  \\
RPS14                            & HLA-DPA1-ARHGDIB                \\
RPS12                            & HLA-B-HLA-C                     \\
OLIG1                            & TYROBP-SRGN                     \\
RPLP0                            & HLA-DRA-LAPTM5                  \\
BCAN                             & HLA-DRB1-ARHGDIB                \\
RPS25                            & CD74-HLA-DRB1                   \\
HLA-DMB                          & FCER1G-SRGN                     \\
\hline
\end{tabular}
\end{table}

\begin{table}[h]
\centering
\caption{Top Spearman BDM for single-cell K27M.}
\begin{tabular}{ll}
\hline
\textbf{Node Perturbation Genes} & \textbf{Link Perturbation Genes} \\
\hline
RPS18                             & MTRNR2L2-MTRNR2L1               \\
ORC4                              & MTRNR2L8-MTRNR2L1               \\
RPL23A                            & MTRNR2L2-MTRNR2L8               \\
EEF1A1                            & MTRNR2L8-MTRNR2L10              \\
PTPRZ1                            & UGDH-AS1-LOC646214              \\
RPL27                             & MTRNR2L1-MTRNR2L10              \\
FABP7                             & LOC643406-LOC646214             \\
ITM2C                             & SHISA9-LOC646214                \\
RPS27                             & MTRNR2L2-MTRNR2L10              \\
RPS11                             & UGDH-AS1-SHISA9                 \\
XIST                              & UGDH-AS1-TMEM212                \\
CA10                              & UGDH-AS1-LOC643406              \\
LOC643406                         & TMEM212-LOC646214               \\
PDGFRA                            & LOC643406-SHISA9                \\
MALAT1                            & UGDH-AS1-ORC4                   \\
CHAD                              & TMEM212-LOC643406               \\
APOD                              & UGDH-AS1-ODF2L                  \\
PCDHGC3                           & TMEM212-SHISA9                  \\
LOC286437                         & LOC646214-ODF2L                 \\
RPL31                             & TMEM212-ODF2L                   \\
RPLP2                             & ORC4-LOC646214                  \\
RPS13                             & LOC643406-ODF2L                 \\
DBI                               & SHISA9-ODF2L                    \\
OLIG1                             & ORC4-SHISA9                     \\
MTRNR2L1                          & UGDH-AS1-ASTN2                  \\
MTRNR2L10                         & LOC643406-LOC286437             \\
SERPINE2                          & ORC4-LOC643406                  \\
CST3                              & RPS6-RPS3                       \\
OPHN1                             & LOC286437-LOC646214             \\
SHISA9                            & TMEM212-ORC4                    \\
LOC646214                         & UGDH-AS1-LOC286437              \\
CCL5                              & RPS3A-RPS6                      \\
SPDYE7P                           & ORC4-ODF2L                      \\
MTRNR2L8                          & RPL3-RPS3A                      \\
NKAIN4                            & ASTN2-LOC646214                 \\
HNRNPA1                           & RPL3-RPL4                       \\
RPL10A                            & LOC286437-ODF2L                 \\
UGDH-AS1                          & ASTN2-SHISA9                    \\
TMEM212                           & RPS6-RPS8                       \\
TRIB2                             & RPS18-RPL18A                    \\
\hline
\end{tabular}
\end{table}

\begin{table}[h]
\centering
\caption{g:Profiler IDHWT GBM PID Single Node BDM Analysis. All terms listed were statistically significant (adjusted p-value $<$ 0.05).}
\begin{tabular}{lp{6.5cm}}  
\hline
\textbf{Term Name} & \textbf{Intersections} \\
\hline
Glial cell differentiation & OLIG1, ID2, VIM, GPM6B, TUBA1A, PTN, CLU \\
Gliogenesis & OLIG1, ID2, VIM, GPM6B, TUBA1A, PTN, CLU \\
Sequestering of actin monomers & TMSB4X, TMSB15A, GSN \\
Oligodendrocyte differentiation & OLIG1, GPM6B, PTN, CLU \\
Extracellular exosome & RPS4X, ANXA2, MEST, RPS11, EEF1A1, RPL26, RPL11, VIM, ACTG1, GSN, TUBA1A, TPI1, CLU, HLA-C, CD63 \\
Extracellular vesicle & RPS4X, ANXA2, MEST, RPS11, EEF1A1, RPL26, RPL11, VIM, ACTG1, GSN, TUBA1A, TPI1, CLU, HLA-C, CD63 \\
Cytoskeleton & TUBB2B, EEF1A1, CNN3, TMSB4X, SLC25A5, TMSB15A, VIM, ACTG1, GSN, TUBA1A, CLU \\
Response of EIF2AK4 (GCN2) to amino acid deficiency & RPS4X, RPS11, RPL26, RPL11, RPL15 \\
Axon guidance & RPS4X, TUBB2B, RPS11, RPL26, RPL11, ACTG1, TUBA1A, RPL15 \\
Nervous system development & RPS4X, TUBB2B, RPS11, RPL26, RPL11, ACTG1, TUBA1A, RPL15 \\
Cellular response to starvation & RPS4X, RPS11, RPL26, RPL11, RPL15 \\
RHO GTPases activate IQGAPs & TUBB2B, ACTG1, TUBA1A \\
Gap junction trafficking and regulation & TUBB2B, ACTG1, TUBA1A \\
pERK-vimentin-KPNA2 complex & KPNA2, VIM \\
\hline
\end{tabular}
\end{table}

\begin{table}[h]
\centering
\caption{g:Profiler K27M PID Single Node BDM Analysis. All terms listed were statistically significant (adjusted p-value $<$ 0.05).}
\begin{tabular}{lp{6.5cm}}  
\hline
\textbf{Term Name} & \textbf{Intersections} \\
\hline
Glial cell development & PLP1, NTRK2, CNTN1, GFAP \\
EIF2AK4 (GCN2) Response & RPS18, RPS11, RPL9, RPL39 \\
Axon guidance & TUBB2A, RPS18, CNTN1, RPS11, RPL9, RPL39 \\
Factor: HSF1; motif: NRGAANNTTCYRGAA & RPS18, MAT2A, CNTN1, CRYAB, RPL39 \\
Nop56p-associated pre-rRNA complex & RPS18, RPS11, RPL9, RPL39 \\
\hline
\end{tabular}
\end{table}

\clearpage
\subsection{Enrichment Signatures from PID-Based Networks and BDM Perturbation}
\label{si:gp_results}

To forecast complex cell fate dynamics, we leveraged AID's graph-theoretic network complexity measures, capturing the causal structure and emergent patterning of cell fate decisions and decode the plasticity signatures steering their goal-directed (teleonomic) behaviors, and aberrant signaling dynamics (i.e., semiotics). Our findings reveal that rather than occupying fixed identities, glioma cells traverse a neurodevelopmental continuum of hybrid states with a differentiation trajectory towards neuronal lineage identities—highlighting fate plasticity as the evolvability engine driving heterogeneity and therapeutic resistance in tumor ecologies. The functional programs and developmental processes underlying these network features (BDM signatures), were further established via gene set enrichment analyses.

g:Profiler enrichment analysis of DEGs within the PID-inferred networks revealed several significant neurodevelopmental and cellular processes in the combined glioma dataset. Notably, the process \textit{forebrain dorsal/ventral pattern formation} (GO:0021798) was enriched with an adjusted \( p \)-value of 3.74 \(\times\) 10\(^{-2}\), and involved genes such as \textit{SFTA3}, \textit{SIX3}, \textit{ISL1}, and \textit{NKX2-1}. Additional enriched terms included \textit{phagocytic vesicle membrane} (GO:0030670, \( p \) = 1.76 \(\times\) 10\(^{-2}\)) and \textit{pre-implantation embryo} (WP:WP3527, \( p \) = 1.08 \(\times\) 10\(^{-2}\)). Several of these gene signatures were independently supported by the single-node BDM perturbation analysis.

Link-level PID perturbation analysis  further highlighted processes associated with neuronal signaling and membrane remodeling. These included \textit{epinephrine binding} (involving \textit{RNLS}), \textit{regulation of short-term synaptic plasticity} (including \textit{CXCR4}, \textit{SYT4}, and \textit{SHISA7}), as well as pathways related to \textit{memory} and \textit{dendritic tree organization}.

In DIPG-specific PID node perturbations, we observed strong enrichment for neurodevelopmental pathways such as \textit{nervous system development} (GO:0007399, \( p \) = 4.88 \(\times\) 10\(^{-4}\)), \textit{pattern specification}, \textit{neurogenesis}, and \textit{forebrain/brain development}. Enrichment of transcription factor motifs further supported these findings, with significant hits for \textit{GKLF} (TF:M01835\_1) and \textit{WT1} (TF:M07436\_1), each with adjusted \( p \)-values $<$ 0.01.

These results reinforce the role of key neurodevelopmental regulators—particularly \textit{FOXG1}, \textit{DLX5}, and \textit{DLX6}—in early brain patterning, neurogenesis, and cancer-specific reprogramming. Many of the glioma-distinguishing gene signatures represent oncofetal morphogens: developmental signals reactivated in cancer that influence both tumor progression and differentiation potential.

Finally, several PID and BDM-identified genes suggest metabolic reprogramming within glioma subtypes. For instance, \textit{FOXD1}, enriched in IDHWT GBM, modulates oxidative stress resistance and mitochondrial function through WNT7A/B signaling interactions \cite{cheng2016foxd1, moparthi2023fox}. In DIPG, transcription factors such as \textit{FOXG1} and \textit{FOXP1} regulate PI3K/AKT, mTOR, and MAPK/ERK pathways, thereby influencing glucose metabolism and energy homeostasis \cite{castaneda2022fox, kaminskiy2022foxp1}. Additional targets such as \textit{IGF1} (a PI3K activator) and \textit{FABP5} (involved in fatty acid uptake and β-oxidation) suggest converging axes of metabolic, transcriptional, and developmental regulation. Together, these findings highlight the multi-layered complexity of glioma phenotypic plasticity, encompassing transcriptional, metabolic, proteomic, epigenetic, and biopsychosocial dimensions of multi-scale processes steering cell fate decisions.

\begin{table}[H]
\small
\setlength{\tabcolsep}{4pt}
\renewcommand{\arraystretch}{1.05}
\centering
\caption{Top GO Biological Process terms enriched in DESeq2 DEGs (combined gliomas vs.\ normal hNSC) using g:Profiler.}
\label{tab:go_glioma_hnsc_top}
\begin{tabularx}{\textwidth}{@{}l p{3.6cm} p{8.6cm} l@{}}
\toprule
\textbf{GO ID} & \textbf{Description} & \textbf{Genes} & \textbf{p.adjust} \\
\midrule
GO:0050804 & modulation of chemical synaptic transmission &
FYN, CACNG5, SLC8A3, GRIK3, IGSF11, ELAVL4, CX3CL1, PCDH17, SHISA9, BCHE, UNC13A, STXBP1, TSHZ3, SLC4A8, DGKI, SLC1A1, SEPTIN5, NLGN4X, NTRK2, RIMS2, PLPPR4, MPP2, CACNG8, SNAP25, CNR1, LRRK2, SHISA6, CSPG5, KCNB1, RIMS1, CACNG7, S100B, ADCY1, DLGAP1, EGR2, PLCL1, NLGN4Y, GRM2, CAMK2B, GRID1, HAP1, MME, MAPK8IP2, NRXN1, DCC, GRIK2, EPHB1, NPAS4, NTNG2, GRID2IP, MAP1A, GRM8, LRRC4C, NPTX1, ROR2, GRIN2A, RIMS4, ITPKA, NPY5R, PTK2B, GRIK4, SLC6A1, SHANK1, SYP, SRGN, CBLN1, NLGN3, RIMS3, CNIH2, CNTN4, CAMK2A, SHANK2, CCL2, AGT, ADORA2A, PRRT2, RELN, NTRK1, CALB1, KCNJ10, GRIN2D, TUBB2B, IL1B, PRKCB, LAMA2, DRD2, ARC, ADRA1A, NRGN, SLC8A2, GRIK5, PTN, SLC4A10, SHISA7, CLSTN2, NEURL1, SYT4, STX1B, JPH4, GFAP, PRKN, GRM3, MAPT & $3.42\!\times\!10^{-19}$ \\
\addlinespace[0.25em]
GO:0099177 & regulation of trans-synaptic signaling &
FYN, CACNG5, SLC8A3, GRIK3, IGSF11, ELAVL4, CX3CL1, PCDH17, SHISA9, BCHE, UNC13A, STXBP1, TSHZ3, SLC4A8, DGKI, SLC1A1, SEPTIN5, NLGN4X, NTRK2, RIMS2, PLPPR4, MPP2, CACNG8, SNAP25, CNR1, LRRK2, SHISA6, CSPG5, KCNB1, RIMS1, CACNG7, S100B, ADCY1, DLGAP1, EGR2, PLCL1, NLGN4Y, GRM2, CAMK2B, GRID1, HAP1, MME, MAPK8IP2, NRXN1, DCC, GRIK2, EPHB1, NPAS4, NTNG2, GRID2IP, MAP1A, GRM8, LRRC4C, NPTX1, ROR2, GRIN2A, RIMS4, ITPKA, NPY5R, PTK2B, GRIK4, SLC6A1, SHANK1, SYP, SRGN, CBLN1, NLGN3, RIMS3, CNIH2, CNTN4, CAMK2A, SHANK2, CCL2, AGT, ADORA2A, PRRT2, RELN, NTRK1, CALB1, KCNJ10, GRIN2D, TUBB2B, IL1B, PRKCB, LAMA2, DRD2, ARC, ADRA1A, NRGN, SLC8A2, GRIK5, PTN, SLC4A10, SHISA7, CLSTN2, NEURL1, SYT4, STX1B, JPH4, GFAP, PRKN, GRM3, MAPT & $3.42\!\times\!10^{-19}$ \\
\addlinespace[0.25em]
\bottomrule
\end{tabularx}
\end{table}

\begin{table}[H]
\small
\setlength{\tabcolsep}{4pt}
\renewcommand{\arraystretch}{1.05}
\centering
\label{tab:go_glioma_hnsc_top}
\begin{tabularx}{\textwidth}{@{}l p{3.6cm} p{8.6cm} l@{}}
\toprule
\textbf{GO ID} & \textbf{Description} & \textbf{Genes} & \textbf{p.adjust} \\
\midrule
GO:0042391 & regulation of membrane potential &
FHL1, SLC8A3, MYC, GRIK3, MYH14, GABRD, IGSF11, GABRA5, KCNK2, CACNA1H, GABRB2, DMD, SLC4A4, PIEZO2, KCNK10, SCN8A, CHRNA4, HCN1, SLC4A8, NLGN4X, NTRK2, RIMS2, MPP2, CNR1, KCND3, PTPN3, LRRK2, KCNB1, SCN2A, RIMS1, KCNQ3, ATP1B2, KCNH2, GABRB3, GABRQ, GRID1, SEZ6, BVES, SCN3A, HCN2, MAPK8IP2, NRXN1, GRIK2, NOS1AP, ACTN2, SCN1A, ABAT, NPAS4, GRIN2A, RIMS4, DSP, PTK2B, GRIK4, SHANK1, SCN4B, CBLN1, NLGN3, RIMS3, CNIH2, TRDN, BOK, CHRNA9, FGF12, ADORA2A, GABRA2, RELN, KCNJ10, GRIN2D, P2RX7, KCNK15, KCNA2, HCN3, IFI6, DRD2, GABRG3, KCNH7, ADRA1A, KCNC1, SLC8A2, CACNA1G, ATP1A3, GRIK5, KCNH5, SLC4A3, KCNMB2, KCNK9, STX1B, NRCAM, RYR2, CTNNA3, GPR35, PRKN, RGS7BP, KCNN2, GABRR1, GLRA3, KCNH8, KCNA1, GABRB1, MAPT & $3.96\!\times\!10^{-18}$ \\
\addlinespace[0.25em]
GO:0050808 & synapse organization &
FYN, PDZRN3, SLC8A3, SLITRK3, CX3CL1, PCDH17, GABRB2, NFASC, UNC13A, PTPRO, NRXN2, PALM, L1CAM, SLC1A1, NLGN4X, PCDHB5, C1QL1, EPHA7, NTRK2, ADGRB3, CTTNBP2, PCDHB9, ADGRL3, GPM6A, FLRT3, LRRC4B, PCDHB10, CNKSR2, IL1RAP, LRRK2, SHISA6, ELFN1, SLIT1, KIRREL3, CTNND2, DSCAM, SLITRK2, GABRB3, NLGN4Y, CAMK2B, SEZ6, LZTS3, ZDHHC15, NGEF, NTN1, LRRN1, BCAN, NRXN1, PCDHB16, SDK2, EPHB1, LRFN5, NOS1AP, SNCB, SPARCL1, NPAS4, NTNG2, LRRC4C, KIF1A, NPTX1, NTRK3, ITPKA, SLC6A1, SHANK1, PCDHB14, SRGN, SYNDIG1, CBLN1, NLGN3, LRRTM3, SHANK2, SRCIN1, GABRA2, RELN, LGI2, NTRK1, DRD2, ARC, IGFN1, TLR2, SLC8A2, SHISA7, CLSTN2, NEURL1, DRP2, EFNA1, VSTM5, NRCAM, LINGO2, LHFPL4, SEMA3E, GDNF, MAPT & $4.99\!\times\!10^{-15}$ \\
\addlinespace[0.25em]
\bottomrule
\end{tabularx}
\end{table}

\begin{table}[H]
\small
\setlength{\tabcolsep}{4pt}
\renewcommand{\arraystretch}{1.05}
\centering
\label{tab:go_glioma_hnsc_top}
\begin{tabularx}{\textwidth}{@{}l p{3.6cm} p{8.6cm} l@{}}
\toprule
\textbf{GO ID} & \textbf{Description} & \textbf{Genes} & \textbf{p.adjust} \\
\midrule
GO:0061564 & axon development &
FYN, FEZ1, DAB1, KALRN, FOXB1, SLITRK3, NFASC, KREMEN1, MAP2, PTPRO, HOXA2, STXBP1, L1CAM, POU3F2, PTPRZ1, MAP6, EPHA7, APLP1, NTRK2, PLPPR4, SLIT2, PAX6, FLRT3, CNR1, NEO1, LAMA1, CHL1, EPHA5, COBL, SEMA6D, CSPG5, SLIT1, S100B, EDN3, NOVA2, DPYSL5, ADCY1, DSCAM, SLITRK2, XK, EGR2, SEMA5B, ATL1, ISL1, NOTCH3, NTN1, NRXN1, DCC, GBX2, EPHB1, APOD, CXCL12, CRMP1, ATP8A2, TSPAN2, NTNG2, MAP1A, OLFM1, VSTM2L, LRRC4C, NPTX1, ANOS1, PLP1, SPTBN4, FGFR2, LAMA3, NLGN3, CNTN4, ARHGAP4, LAMC2, RET, RTN4RL2, RELN, NTRK1, RTN4R, NKX6-1, EPHB6, MT3, TUBB2B, CRTAC1, LAMA2, DRD2, SLIT3, PTN, RGMA, FN1, NOG, EFNA1, TNN, NRCAM, LHX2, EFNA2, RNF165, SEMA6A, SEMA3E, GDNF, EPHA6, CNTN6, MAPT & $5.55\!\times\!10^{-14}$ \\
\addlinespace[0.25em]
GO:0007409 & axonogenesis &
FYN, FEZ1, DAB1, KALRN, FOXB1, SLITRK3, NFASC, MAP2, PTPRO, HOXA2, STXBP1, L1CAM, POU3F2, PTPRZ1, MAP6, EPHA7, APLP1, NTRK2, PLPPR4, SLIT2, PAX6, FLRT3, NEO1, LAMA1, CHL1, EPHA5, COBL, SEMA6D, SLIT1, S100B, EDN3, NOVA2, DPYSL5, ADCY1, DSCAM, SLITRK2, XK, EGR2, SEMA5B, ATL1, ISL1, NOTCH3, NTN1, NRXN1, DCC, GBX2, EPHB1, CXCL12, CRMP1, ATP8A2, NTNG2, MAP1A, OLFM1, VSTM2L, LRRC4C, NPTX1, ANOS1, SPTBN4, FGFR2, LAMA3, NLGN3, CNTN4, ARHGAP4, LAMC2, RET, RELN, NTRK1, RTN4R, NKX6-1, EPHB6, MT3, LAMA2, DRD2, SLIT3, FN1, NOG, EFNA1, TNN, NRCAM, LHX2, EFNA2, RNF165, SEMA6A, SEMA3E, GDNF, EPHA6, CNTN6, MAPT & $1.68\!\times\!10^{-12}$ \\
\bottomrule
\end{tabularx}
\end{table}

\end{document}